# Two-dimensional electronic spectroscopy of bacteriochlorophyll *a* with synchronized dual mode-locked lasers

JunWoo Kim,[1,†] Jonggu Jeon,[1] Tai Hyun Yoon,[1,2*] and Minhaeng Cho[1,3*]

[1]*Center for Molecular Spectroscopy and Dynamics, Institute for Basic Science (IBS), Seoul 02841, Republic of Korea*

[2]*Department of Physics, Korea University, Seoul 02841, Republic of Korea*

[3]*Deparment of Chemistry, Korea University, Seoul 02841, Republic of Korea*

*Electronic mails: mcho@korea.ac.kr (MC) or thyoon@korea.ac.kr (THY)

[†]*Present address: Department of Chemistry, Princeton University, Princeton, NJ 08544, USA*

**Abstract**

**How atoms and electrons in a molecule move during a chemical reaction and how rapidly energy is transferred to or from the surroundings can be studied with flashes of laser light. However, despite prolonged efforts to develop various coherent spectroscopic techniques, the lack of an all-encompassing method capable of both femtosecond time resolution and nanosecond relaxation measurement has hampered various applications of studying correlated electron dynamics and vibrational coherences in functional materials and biological systems. Here, we demonstrate that two broadband (>300 nm) synchronized mode-locked lasers enable two-dimensional electronic spectroscopy (2DES) study of chromophores such as bacteriochlorophyll *a* in condensed phases to measure both high-resolution coherent vibrational spectrum and nanosecond electronic relaxation. We thus anticipate that the dual mode-locked laser-based 2DES developed and demonstrated here would be of use for unveiling the correlation between the quantum coherence and exciton dynamics in light-harvesting protein complexes and semiconducting materials.**

**Introduction**

Our understanding of natural light-harvesting systems involving excitonic energy transfers has been revolutionized by the application of femtosecond nonlinear optical spectroscopy.[1,2] Delocalized quantum excited states created by photoexcitation of and electronic couplings between light-absorbing chromophores relax through a myriad of pathways, such as exciton migration, excitation localization induced by the thermal fluctuation of coupled bath degrees of freedom, and long-distance Förster energy transfer.[3-5] A series of events involving electron and proton transfer reactions subsequently occur in photosynthetic systems. One of the most crucial techniques for studying such complex photoinduced reaction dynamics of generic self-assembled multichromophoric systems is 2DES,[1,6-8] which is capable of providing the time correlation between the initial and final states by mapping their nonlinear optical response onto 2D excitation- and detection-frequency space.[9,10]

The 2DES requires multiple femtosecond optical pulses to interrogate molecular systems, such as light-harvesting protein complexes,[1,11-14] optical chromophores in solutions,[10,15] semiconducting materials,[16,17] metallic nanoparticles,[18] and chiral aggregates[19]. It provides information on not only ultrafast energy transfer kinetics but also dephasing time scales of electronic and/or vibrational coherences of coupled chromophore systems like photosynthetic proteins.[12,20] Recently, the 2DES of Fenna-Matthew-Olson complex revealed an important role of vibronic couplings in the ultrafast energy transfer process on the one-exciton state manifold.[6,8,13] It turns out that the vibronically excited vibrational modes play a crucial role in enhancing the pigment-to-pigment energy transfer rates.[21,22] In addition, the excited-state dynamics and relaxations of chlorophyll molecules that are critical pigments in photosynthetic complexes are of fundamental importance.[23-25]

In 2DES experiments, the time evolution of excited states initiated by a pair of pump pulses generating temporally interfered electric fields is monitored by using another time-delayed probe and local oscillator pulses. The waiting time ($T_w$) that is the time delay between the pump and probe pulses is usually scanned by mechanically changing the optical path lengths of the excitation and detection pulses. It was shown that the $T_w$-scanning speed can be dramatically enhanced by utilizing two repetition rate-stabilized mode-locked lasers (MLs), where the technique is called asynchronous optical sampling (ASOPS).[26] In ASOPS, each ML generates a train of optical pulses separated by an equal time interval. When the repetition rates of the two MLs are slightly detuned as $f_r$ and $f_r + \Delta f_r$, the optical time delay ($T_w$) between a pair of pulses from the two MLs increases linearly in laboratory time ($t$). The corresponding time delay increment ($\Delta T_w$) is determined by the repetition rate and the detuning factor $\Delta f_r$ as

$$\Delta T_w = \frac{1}{f_r} - \frac{1}{f_r + \Delta f_r} \cong \frac{\Delta f_r}{f_r^2} = f_D \frac{1}{f_r}. \quad (1)$$

Here, $f_D$ is the down-conversion factor defined by $\Delta f_r / f_r$. Although the automatic time delay scanning scheme referred to as ASOPS was proposed and demonstrated in 1987,[26] the application of ASOPS to coherent multidimensional spectroscopy has not been achieved, even though the ASOPS has been employed in the development of dual-frequency comb spectroscopic techniques.[27,28] Recently, synchronized MLs without photodetector array nor mechanical time-delay devices have been used to perform Raman microspectroscopy[29,30], dual-frequency comb 2DES[31], and time- and frequency-resolved transient absorption and refraction spectroscopy[32-34].

Here, we experimentally demonstrate that synchronized mode-locked laser-based 2DES (SM-2DES) without a long mechanical delay line for the $T_w$–scan enables ultrafast time-resolved studies of sub-picosecond electronic and vibrational dephasings, picosecond solvation dynamics, and nanosecond population relaxation without sacrificing its femtosecond time-resolvability and detection sensitivity. We show that the data acquisition time for the measurement of the $T_w$-dependent 2DES signals is substantially shortened in comparison to the conventional 2DES technique. We, here, present the SM-2DES experimental results for bacteriochlorophyll *a* (BChla)/1-propanol solution to demonstrate the feasibility and capability of a novel coherent 2D spectroscopic technique. It should be noted that the ASOPS-based femtosecond time-resolved spectroscopy with an interferometric detection of the nonlinear signal[34], including SM-2DES, requires tightly repetition-rate-stabilized MLs to achieve a sub-fs timing jitter. The fast and efficient data acquisition rate of SM-2DES enabled us to study ultrafast solvation dynamics, vibrational coherences of vibronically coupled modes, and non-Condon effects on the coherent vibrational spectra extracted from the SM-2DES signals.

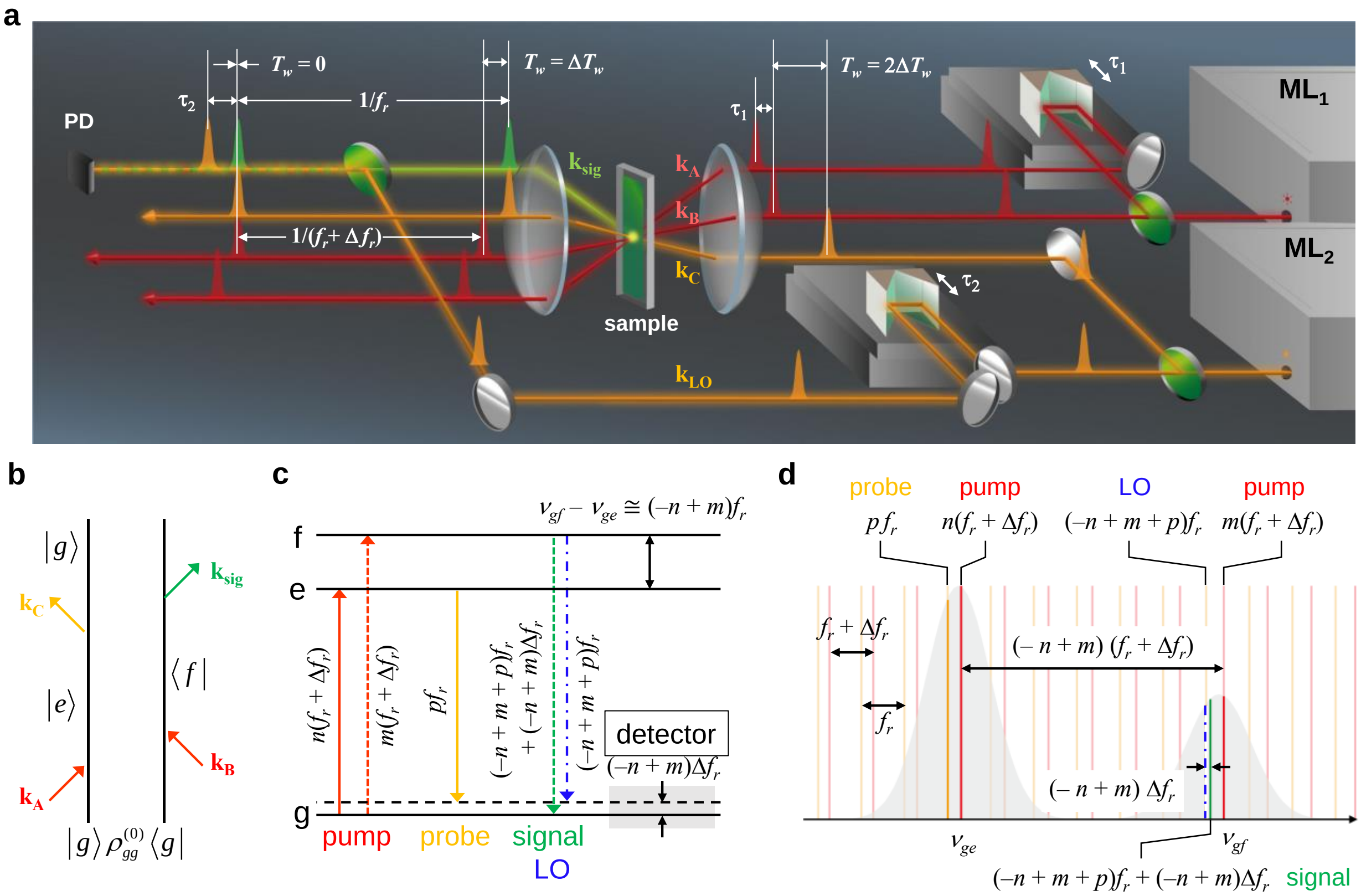

**Fig. 1 | Synchronized mode-locked laser-based two-dimensional electronic spectroscopy. a**. Schematic representation of the SM-2DES experimental setup. Green circular plates represent beam splitters. The $ML_1$ beam is split into two beams with wave vectors $\mathbf{k}_A$ and $\mathbf{k}_B$, and the time delay $\tau_1$ between the two pulsed beams is scanned using a translational stage. The pulses propagating along the directions determined by $\mathbf{k}_A$ and $\mathbf{k}_B$ are used as a pair of pump pulses to excite optical chromophores in solution. The $ML_2$ beam is split into two beams with wave vectors $\mathbf{k}_C$ and $\mathbf{k}_{LO}$. The time delay $\tau_2$ between the $\mathbf{k}_C$ and $\mathbf{k}_{LO}$ pulses is scanned using another translational stage. The transient grating generated by $\mathbf{k}_A$ and $\mathbf{k}_B$ beams diffracts the probe with wave vector $\mathbf{k}_C$ into the direction satisfying the desired phase-patching condition $\mathbf{k}_{sig} = -\mathbf{k}_A$ ($ML_1$) + $\mathbf{k}_B$ ($ML_1$) + $\mathbf{k}_C$ ($ML_2$), where the 2DES signal under detection propagates along the direction determined by $\mathbf{k}_{sig}$. The repetition frequencies of the pump and probe beams are $f_r + \Delta f_r$ and $f_r$, respectively. The 2DES signal is combined with the local oscillator field with a wave vector $\mathbf{k}_{LO}$ that is collinear with the propagation direction of the 2DES signal electric field, and the interference signal is detected by a photodetector (PD). **b.** Double-sided Feynman diagram describing one of the various terms contributing to the 2DES signal. **c.** Energy-level diagram of a three-level system. $|g\rangle$ is the ground state, and $|e\rangle$ and $|f\rangle$ represent two vibrational states in the electronically excited state. $n$, $m$, and $p$ are the relevant frequency comb mode numbers. $\nu_{ge}$ ($\nu_{gf}$) indicates the optical transition frequency between $|g\rangle$ and $|e\rangle$ ($|f\rangle$) states. The vibrational quantum beat with a frequency of $\nu_{gf} - \nu_{fe} \approx (-n+m) f_r$ is measured through the interferometric detection of the 2DES signal with LO from $ML_2$. The beat signal contributing to the time-domain interferogram oscillates with a frequency of $(-n+m)\Delta f_r$, which is in the RF domain. **d.** Comb structures of the pump, probe, and LO fields in the frequency domain. The line-broadened absorption bands of the $|g\rangle$-$|e\rangle$ and $|g\rangle$-$|f\rangle$ transitions are shown in grey.

## Results

**Synchronized mode-locked laser-based 2D electronic spectroscopy.** The experimental setup for SM-2DES is illustrated in Fig. 1**a** (see Supplementary Note 1 for detailed optical layout). SM-2DES in a boxcar geometry requires four temporally and spatially separated pulses that are referred to as their wave vectors $\mathbf{k}_A$, $\mathbf{k}_B$, $\mathbf{k}_C$, and $\mathbf{k}_{LO}$, where LO represents the local oscillator used to interferometrically detect the generated four-wave-mixing signal electric field.[10] The repetition rates of $ML_1$ and $ML_2$ are slightly different from each other by $\Delta f_r$ (Fig. 1**a**), where the repetition rate ($f_r$) of $ML_2$ is 80 MHz, and that of $ML_1$ is 80 MHz + $\Delta f_r$. $\Delta f_r$ was chosen as 38.4 Hz and 3.2 kHz to set $\Delta T_w$ = 6 fs and 500 fs, respectively, according to equation (1). First, two pump pulses with wave vectors $\mathbf{k}_A$ and $\mathbf{k}_B$, respectively, that are used to excite chromophores in solutions are produced by $ML_1$, and the time-delay between the two pulses, denoted as $\tau_1$, is experimentally controlled using a high-precision translational stage. This spatial interference fringe generated by the two pump fields propagating in the two different directions determined by the wave vectors of $\mathbf{k}_A$ and $\mathbf{k}_B$ acts like a grating. Then, the probe pulse propagating along the direction of $\mathbf{k}_C$ arrives at the sample after a waiting time $T_w$, and then the probe beam is scattered due to the third-order field-matter interaction. Such a three-pulse scattering process was initially studied by Weiner and Ippen.[35] The $T_w$-dependent relaxation of the transient grating diffracted in the direction of $\mathbf{k}_{sig}$ (= $-\mathbf{k}_A + \mathbf{k}_B + \mathbf{k}_C$) results from the decay of the population of excited-state molecules in a dissipative medium. The scattered signal is under interferometric detection with an added LO pulse (with wave vector $\mathbf{k}_{LO} = \mathbf{k}_{sig}$) from $ML_2$. Here, the delay time between the probe and LO pulses, denoted as $\tau_2$, is scanned using another high-precision translational stage.

Figure 1**b** depicts one of the double-sided Feynman diagrams[36,37] describing a photon-echo-process contributing to the 2DES signal, where $|g\rangle$ is the electronic ground state and $|e\rangle$ and $|f\rangle$

denote two different vibrational states in the electronically excited state. Because the signal field is interferometrically detected with LO field from $ML_2$, the measured 2DES signal becomes independent of carrier-envelope phase slips of the two MLs.[38] As can be seen in the energy level diagram and comb structures of the pulse spectra in Figs. 1**c** and 1**d**, respectively, the vibrational coherence with a frequency of $\nu_{gf} - \nu_{fe} \approx (-n+m)f_r$ can be measured by allowing the third-order 2DES signal field to interfere with a specific LO comb mode, which results in a beat signal with an RF of $(-n+m)\Delta f_r$. Here, $n$, $m$, and $p$ are the frequency comb mode numbers.

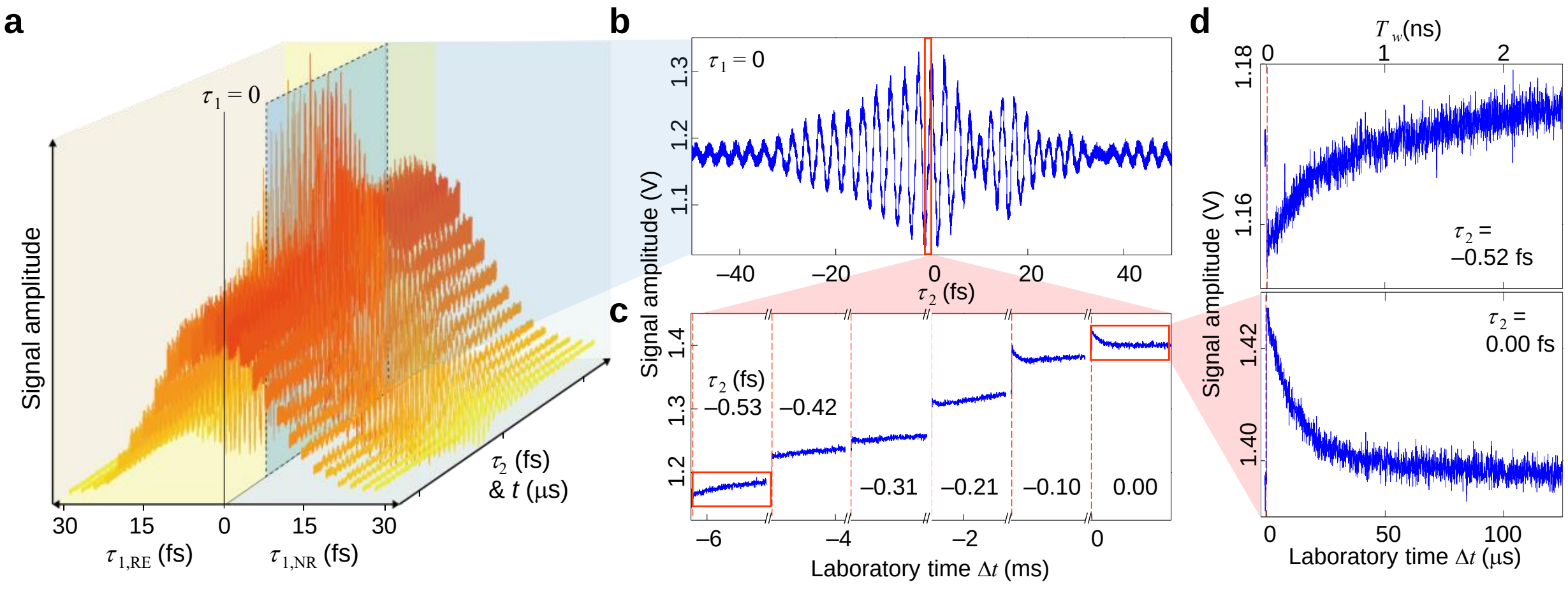

**Fig. 2 | Time-domain interferometric 2DES signal. a**. Experimentally measured SM-2DES raw data of IR125/ethanol solution with respect to $\tau_1$ and $\tau_2$. Here, the repetition frequency detuning factor $\Delta f_r$ is 1.6 kHz. **b.** SM-2DES interference signal with respect to $\tau_2$ when $\tau_1$ = 0 fs. The $T_w$-dependent decaying signals are hidden in the prominent $\tau_2$-dependent interference signal. **c.** Zoomed-in signal taken from the red box in **b**. **d.** Zoomed-in $T_w$-dependent signals $\tau_2$ = –0.52 fs (top) and $\tau_2$ = 0.00 fs (bottom) at $\tau_1$ = 0 fs.

Figure 2**a** shows a representative set of SM-2DES raw data of the IR125/ethanol solution. When the pump pulse with wave vector $\mathbf{k}_A$ precedes the $\mathbf{k}_B$ pulse, the photon echo response dominates the detected 2DES signal for an optically heterogeneous system. The opposite time-ordered field-matter interactions in the order of $\mathbf{k}_B \rightarrow \mathbf{k}_A \rightarrow \mathbf{k}_C$ pulses generate the non-rephasing 2DES signal. At a fixed $\tau_1$, a continuous $\tau_2$-scan is performed while ASOPS with two trains of $\mathbf{k}_B$ and $\mathbf{k}_C$ pulses from the two MLs is used to scan the waiting time ($T_w$) that is related to laboratory

time $t$ by $T_w = f_D \Delta t$. $\Delta t$ is the time elapsed after the optical trigger signal generated when a pair of pulses from the two MLs temporally overlap. The most prominent interference pattern of the SM-2DES signal at $\tau_1 = 0$ shown in Fig. 2**b** arises from the interference between the local oscillator electric field and the 2DES signal electric field. However, the interferogram in Fig. 2**b** depends on $T_w$ too, even though it is not clearly visible in the interferogram plotted in Fig. 2**b**. One can retrieve the $T_w$-dependent 2DES signals by rearranging the raw data. The $T_w$ -dependent feature of the SM-2DES signal can be clearly seen in the zoomed-in plot in Fig. 2**c**. Each decay curve (Fig. 2**d**) at fixed $\tau_1$ and $\tau_2$ corresponds to the measurement (or waiting) time-dependent interference signal, i.e., $I(\tau_2, \Delta t, \tau_1) = I(\tau_2, T_w / f_D, \tau_1)$, which shows how the vibrational quantum beats in the infrared frequency domain can be measured with a photodetector in the RF domain by employing an ASOPS-based down-conversion measurement scheme. The experimental data in Fig. 2**a** is, therefore, a 3D time-domain interferometric signal $I(\tau_2, \Delta t, \tau_1)$ that depends on $\tau_1$, $\Delta t$ ($= T_w/f_D$), and $\tau_2$.

To establish the relationship between the measured time-domain interferogram in Fig. 1**b** and the nonlinear response function $S^{(3)}(t_3, t_2, t_1)$ of optical chromophores in condensed phases, we employ time-dependent perturbation theory[37,39], use optical frequency comb field expressions[38], and invoke an impulsive pulse approximation. One can find that the experimentally measured signal $I(\tau_2, \Delta t, \tau_1)$ that results from the interference between the LO and 2DES signal electric fields, $\mathbf{E}_{LO}$ and $\mathbf{E}_{signal}$, respectively, is given by (see Supplementary Note 2 for detailed theoretical derivation and ref [40]):

$$I(\tau_2, \Delta t, \tau_1) \propto \mathrm{Re}\left[\mathbf{E}_{LO}^*(t) \cdot \mathbf{E}_{\mathrm{signal}}^*(t)\right] \propto \mathrm{Re}[R_{echo}(\tau_2, T_w, \tau_1)], \quad (2)$$

where $R_{echo}(\tau_2, T_w, \tau_1)$ is the time-domain nonlinear response function associated with photon-echo spectroscopy.[41] Equation (2) suggests that the all-time-resolved SM-2DES interferogram is simply proportional to the real part of the photon-echo response function. As manifested in Eq. (2), the time scale of femtosecond and nanosecond molecular processes during $T_w$ is converted to that of laboratory time $\Delta t$ on the nanosecond to millisecond time scales by the inverse of the down-conversion factor, i.e., $1/f_D = 2.5\times10^4$ and $2.08\times10^6$ for $\Delta f_r = 3.2$ kHz and 38.4 Hz, respectively, with $f_r = 80$ MHz in the present work.

Fourier-transforming the 3D interferometric signal, $I(\tau_2, \Delta t, \tau_1)$, with respect to $\tau_1$ and $\tau_2$, we were able to obtain the 2DES spectra with respect to the excitation angular frequency ($\omega_1$) and detection angular frequency ($\omega_2$). A notable advantage of SM-2DES compared to the conventional 2DES is that both the amplitude and phase of the 2DES data can be analytically normalized and corrected, respectively, by using the interferometric pump-probe signals measured with the same instrument (see Methods and Supplementary Note 3). Here, the phase stability between the probe and LO pulses is maintained by employing a continuous $\tau_2$-scan method[34] that is combined with an additional reference interferometer (see Methods). In practice, it should be noted that the present phase correction and spectral amplitude calibration methods are useful when the complex (real and imaginary) pump-probe data are available. However, the spectral shapes of the pump pulses are unavoidably imprinted in the measured SM-2DES spectra. Since the automatic $T_w$-scan was performed up to 3.0 ns with a time interval of 6 fs or 500 fs, when the repetition rate detuning factor is 38.4 Hz or 3.2 kHz, respectively, the sheer size and information density of 2DES data (many thousands to millions of 2DES spectra) are extraordinarily large, which attests to the capability of our SM-2DES technique.

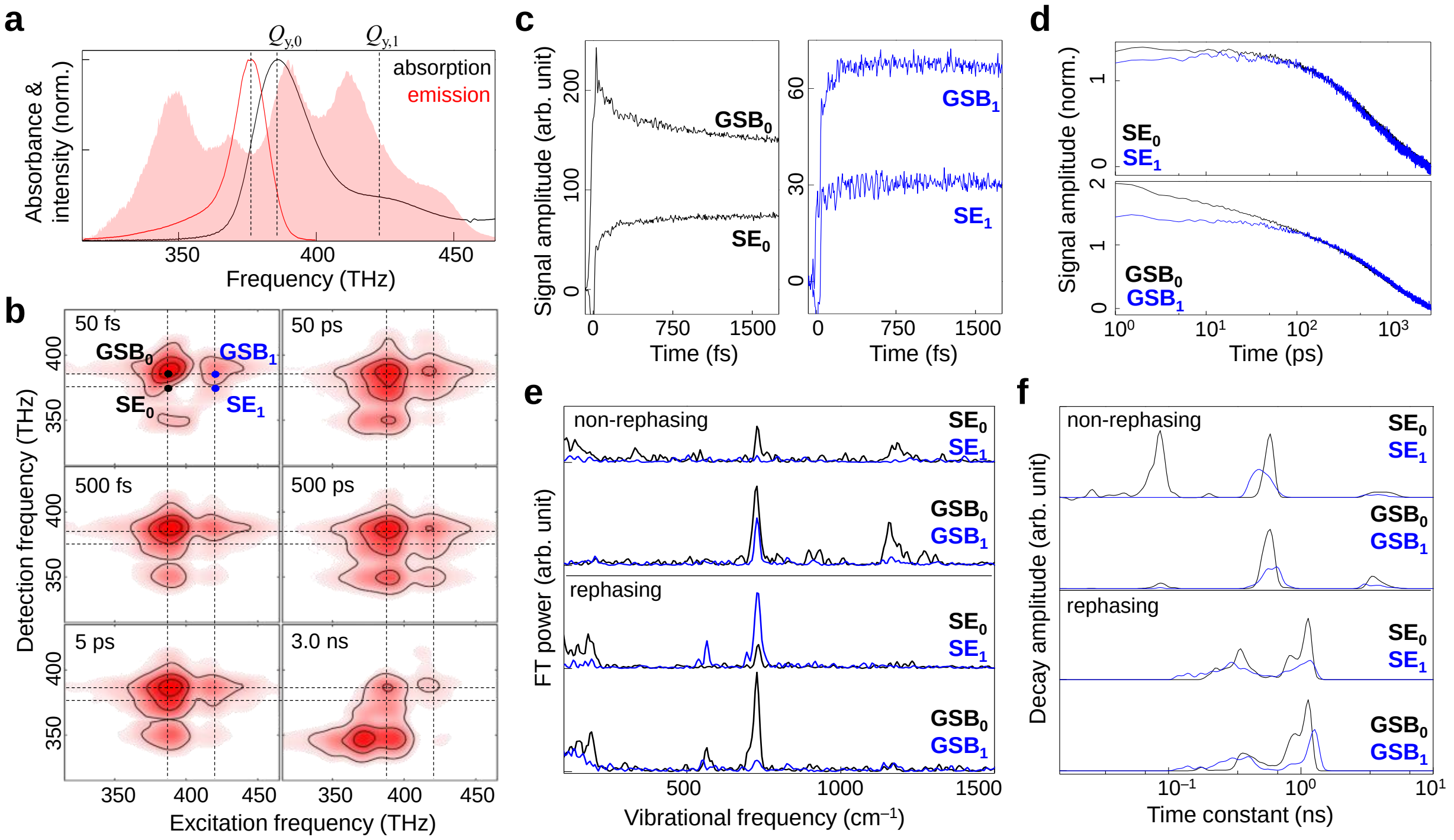

**Fig. 3 | Time-resolved SM-2DES spectra of BChla. a.** The absorption (black) and emission (red) spectra of BChla/1-propanol solution. The power spectrum of the light source (pink area) is presented for the sake of direct comparison. **b.** Absorptive (rephasing + non-rephasing) SM-2DES spectra of the BChla solution at $T_w$ = 50, 500 fs, 5 ps ($\Delta f_r$ = 38.4 Hz), 50 ps, 500 ps and 3 ns ($\Delta f_r$ = 3.2 kHz). Each spectrum is normalized by its maximum amplitude. **c-d.** The time traces of the rephasing 2DES signals at four different positions in the SM-2DES spectra, which are **$SE_0$** (387 THz, 375 THz), **$GSB_0$** (387 THz, 387 THz), **$SE_1$** (420 THz, 375 THz), and **$GSB_1$** (420 THz, 387 THz). **e**. Fourier transforms of the oscillating components in the time traces plotted in **c**. **f.** The distribution of decay time constants associated with the time traces shown in **d**.

**SM-2DES of bacteriochlorophyll *a*.** The spectral bandwidth of our Ti:Sapphire oscillators is broad enough to cover both the $Q_y$-absorption and -emission bands of the BChla in propanol as well as the high-lying vibronic bands (Fig. 3**a**). Three representative transitions between the ground ($S_0$) and $Q_y$ states, which are the 0-0 $Q_y$ absorption ($Q_{y,0}$) at 387 THz, 0-1 $Q_y$-absorption ($Q_{y,1}$) at 420 THz, and 0-0 $Q_y$-emission at 375 THz, are marked by dashed lines. The $Q_{y,0}$-emission peak is Stokes-shifted from the $Q_{y,0}$-absorption peak due to free energy minimization processes such as solvation. The time resolution of our SM-2DES is 12.5 fs (see Method and Supplementary Note 1) so that one can study the wave packet oscillations of vibronically coupled modes with frequencies up to 2670 $cm^{-1}$. During the time required for a full $T_w$–scan, which is $1/\Delta f_r$, the number of $T_w$-points in the dynamic range of the $T_w$-resolved measurements from 0 to 12.5 ns is $2.08\times10^6$ when $\Delta f_r$ = 38.4 Hz and $f_r$ = 80 MHz and $2.5\times10^4$ when $\Delta f_r$ = 3.2 kHz and $f_r$ = 80 MHz. However, the conventional 2DES with a mechanical delay line for the $T_w$–scan is, in practice, not useful to collect such a large number of $T_w$-dependent 2DES signals. Due to the high data acquisition rate of the SM-2DES compared to the conventional 2DES methods using a single mode-locked laser and a mechanical delay line for scanning the waiting time, the frequency resolution of the coherent vibrational spectrum extracted from the time-resolved 2DES signals can be substantially enhanced. Furthermore, the large number of data equally spaced over a wide range of waiting time enables to perform boxcar averaging, which can, in turn, enhance the signal-to-noise ratio of the 2DES spectrum even at a long waiting time at which the amplitude of the signal is expected to be small.

In Fig. 3**b**, the six representative 2DES spectra at waiting times of 50 fs, 500 fs, 5 ps, 50 ps, 500 ps, and 3.0 ns are shown. At a short waiting time ($T_w$ = 50 fs), both ground-state bleach (GSB) and stimulated emission (SE) positively contribute to the absorptive 2DES signal. In general, the coherent electronic artifact significantly contributes to the 2DES signal at zero waiting time when the three pulses that interact with the sample overlap in time.[38] The diagonal peak at (387 THz, 387 THz) and the cross-peak at (420 THz, 387 THz) originating from both the GSB and SE terms become vertically elongated along the detection frequency due to rapid intramolecular vibrational redistribution and solvation dynamics, which causes the detection frequency of the SE contribution to shift down to 375 THz in less than 5 ps.[37] The two GSB contributions denoted as $GSB_0$ at (387 THz, 387 THz) and $GSB_1$ at (420 THz, 387 THz) result from the ground-state bleaches produced at $Q_{y,0}$ by the $Q_{y,0}$- and $Q_{y,1}$-excitations, respectively. The $Q_{y,1}$ diagonal peak at (420 THz, 420 THz) is not clearly visible in Fig. 3**b** because its amplitude is much smaller than the $Q_{y,0}$ diagonal peak, denoted as $GSB_0$, by a factor of approximately 0.06. Also, there should be another GSB peak at (387 THz, 420 THz), which results from the ground-state bleach of $Q_{y,1}$ state when the pump induced a $Q_{y,0}$ transition. The amplitude of the cross-peak at (387 THz, 420 THz) could be comparable to that of $GSB_1$. However, the $GSB_1$ peak is found to be approximately four times stronger than that of the cross-peak at (387 THz, 420 THz) (Supplementary Figure 2c). Such a reduction in the amplitude of the cross-peak at (387 THz, 420 THz) is caused by the negative contribution from an excited-state absorption.[25] The two SE contributions denoted as $SE_0$ at (387 THz, 375 THz) and $SE_1$ at (420 THz, 375 THz) result from the Stokes-shifted excited states $Q_{y,0}$ and $Q_{y,1}$, respectively. The $GSB_1$ and $SE_1$ in Fig. 3b are frequency-resolved from the $GSB_0$ and $SE_0$ peaks (Fig. 3a), even though both the absorption and emission spectra have broad lineshapes with a shoulder peak in the absorption spectrum that originates from a vibronic progression. Such clear separations of the $GSB_1$ and $SE_1$ peaks from the $GSB_0$ and $SE_0$ peaks in the frequency domain are partly due to the non-Gaussian spectral shapes of the pump pulses (Fig. 3a) that affect the lineshape of the 2DES spectrum along the excitation frequency axis. Note that, in general, both the pump and probe spectra affect the spectral features of the 2DES spectrum along the excitation and detection frequencies. Here, we find a positive peak at (387 THz, 350 THz) whose amplitude increases on a time scale of nanoseconds, which could be assigned to the thermal grating contribution that results from nonradiative decay of the excited BChla molecules.

The SM-2DES spectra at long waiting times obtained by employing a fast $T_w$-scan mode ($\Delta f_r$ = 3.2 kHz) (see the 2DES spectra at $T_w$ = 50 ps, 500 ps, and 3.0 ns in Fig. 3**b**) provide information on slow population relaxation and local heating dynamics. The 2D spectral shapes of the four peaks in the 50-ps 2DES spectrum remain constant up to a waiting time of 500 ps, and the corresponding signals decay to zero on a few ns time scale. In the 2DES spectrum at $T_w$ = 3.0 ns, a horizontally elongated positive peak appears, which can be assigned to the formation of a transient thermal grating in the solution sample due to the non-collinearly propagating trains of pulses (see Supplementary Note 4). The slow relaxation and evolution of the thermal grating provide useful information on thermal diffusion processes in the surroundings of chromophores in condensed phases.

**Picosecond relaxation and solvation dynamics.** The time profiles of the 2DES signals at the four peak positions ($GSB_0$, $SE_0$, $GSB_1$, and $SE_1$) over the broad range (six decades) of waiting times are plotted in Figs. 3**c** and 3**d**. The amplitudes of the slow time profiles from 200 ps to 3 ns are matched with each other so that one can directly compare the early parts of the four decaying signals with $T_w$ (Fig. 3**d**). At short waiting times (<100 ps), the decay profiles of $SE_0$ and $GSB_0$ that are associated with the $Q_{y,0}$ excitation differ from those of the $SE_1$ and $GSB_1$ signals, respectively (Figs. 3**c** and 3**d**). Recently, Moca *et al*. carried out a series of 2DES experiments for chlorophyll *a* in a few different solvents[42], and they found that the change of the 2DES spectra at short waiting times is dictated by the solvation dynamics. If the dipole moments of the ground and excited states differ from each other, the surrounding solvent molecules right after an instantaneous electronic excitation of a chromophore are in a non-equilibrium state in terms of solvation. Then, the solvent molecules would rearrange their orientations and move longitudinally from the excited state molecule to stabilize the newly created charge distributions of the solute molecule. Due to the ultrafast solvation dynamics, the SE peak that overlaps with the corresponding GSB peak at $T_w$=0 undergoes a frequency red-shift along the detection frequency axis[37], which makes the diagonal peak vertically broadened and elongated along the detection frequency axis. Therefore, the amplitude of the $GSB_0$ ($SE_0$) peak at (387 THz, 387 THz) ((387 THz, 375 THz)) shows a short-time decaying (rising) feature. Therefore, our observation that the short time changes of the $SE_0$ and $GSB_0$ signals are notably large suggests that the charge distribution of the $Q_{y,0}$ state differs from that of the ground state, $S_0$. This result is consistent with Moca *et al.*'s experimental observation.[42] To confirm this interpretation, we carried out density functional theory (DFT) and time-dependent DFT calculations of the ground ($S_0$) and excited ($Q_y$) states of BChla, respectively. Indeed, these *ab initio* calculation results show that the dipole moments of these two states are distinctively different from each other (Supplementary Figure 4). However, still, it is not entirely clear why the signals at $SE_1$ and $GSB_1$ generated by the transition to the $Q_{y,1}$ state do not relax with $T_w$ for approximately 100 ps. One possible explanation is that the charge distribution of the $Q_{y,1}$ state is similar to that of the ground state, which makes the solvation dynamics contributions to the decays of the $SE_1$ and $GSB_1$ signals small.

**Femtosecond dynamics and vibrational coherences.** Using the short-time 2DES signals at four different points in a given 2DES spectrum, one can obtain the spectra of vibronically excited modes, which will be referred to as coherent vibrational spectrum (CVS) (Fig. 3**e**). We found four strongly vibronically coupled modes with frequencies of 190, 564, 728, and 1156 $cm^{-1}$. Here, the frequency resolution is estimated to be 8 $cm^{-1}$ by considering the scan range of the waiting time ($0 \leq T_w \leq 5.0$ ps) and the width (2.5 ps) of the apodization window. It should be emphasized that the SM-2DES technique does not need additional data acquisition time for a longer waiting time scanning to achieve an improved frequency resolution for the spectral analysis of vibrational quantum beats.

The vibrational mode at 728 $cm^{-1}$ is found to be the most strongly vibronically coupled to the $Q_y$ transition, which indicates that the displacement between the corresponding vibrational potential energy surfaces of the $S_0$ and $Q_y$ states is large. The vibronic coupling between the $Q_y$ transition and the 1156 $cm^{-1}$ mode is responsible for the $Q_{y,1}$ band because the frequency difference between the $Q_{y,0}$ and $Q_{y,1}$ bands in the absorption spectrum is close to 1156 cm$^{-1}$. The amplitudes of the other two modes at 190 and 564 cm$^{-1}$ depend on the excitation and detection frequencies. Although the information density of the three-dimensional 2DES data obtained with wide dynamic range 2DES measurements is very high, it is still difficult to thoroughly analyse the vibrational coherence spectra in Fig. 3**e**. Below, we shall present principal components analysis results and compare the DFT results for the ground and excited states of BChla.

**Nanosecond dynamics.** In Fig. 3**f**, we present the decay-constant distributions of the 2DES signals at the four peak positions, $GSB_0$, $SE_0$, $GSB_1$, and $SE_1$, which are the weighted histograms of the decay constants from the tri-exponential fitting analyses of the waiting-time dependent signals. The amplitude of each exponential component is used as the weighting factor. From the total absorptive 2DES signals (Supplementary Figure 5), we found that the lifetime of the $Q_y$ state is about 2.1 ns, which is consistent with the fluorescence lifetime of BChla in 1-propanol.[43] The sub-nanosecond components could be associated with the rotational relaxation of BChla because such a component was not observed in the magic-angle transient absorption study.[44] Although the nanosecond dynamics of the BChla solution is mainly determined by the rotational diffusion and population relaxation, the SM-2DES with nanosecond scanning capability could be useful for studying systems requiring both 2DES and nanosecond measurements in the future.

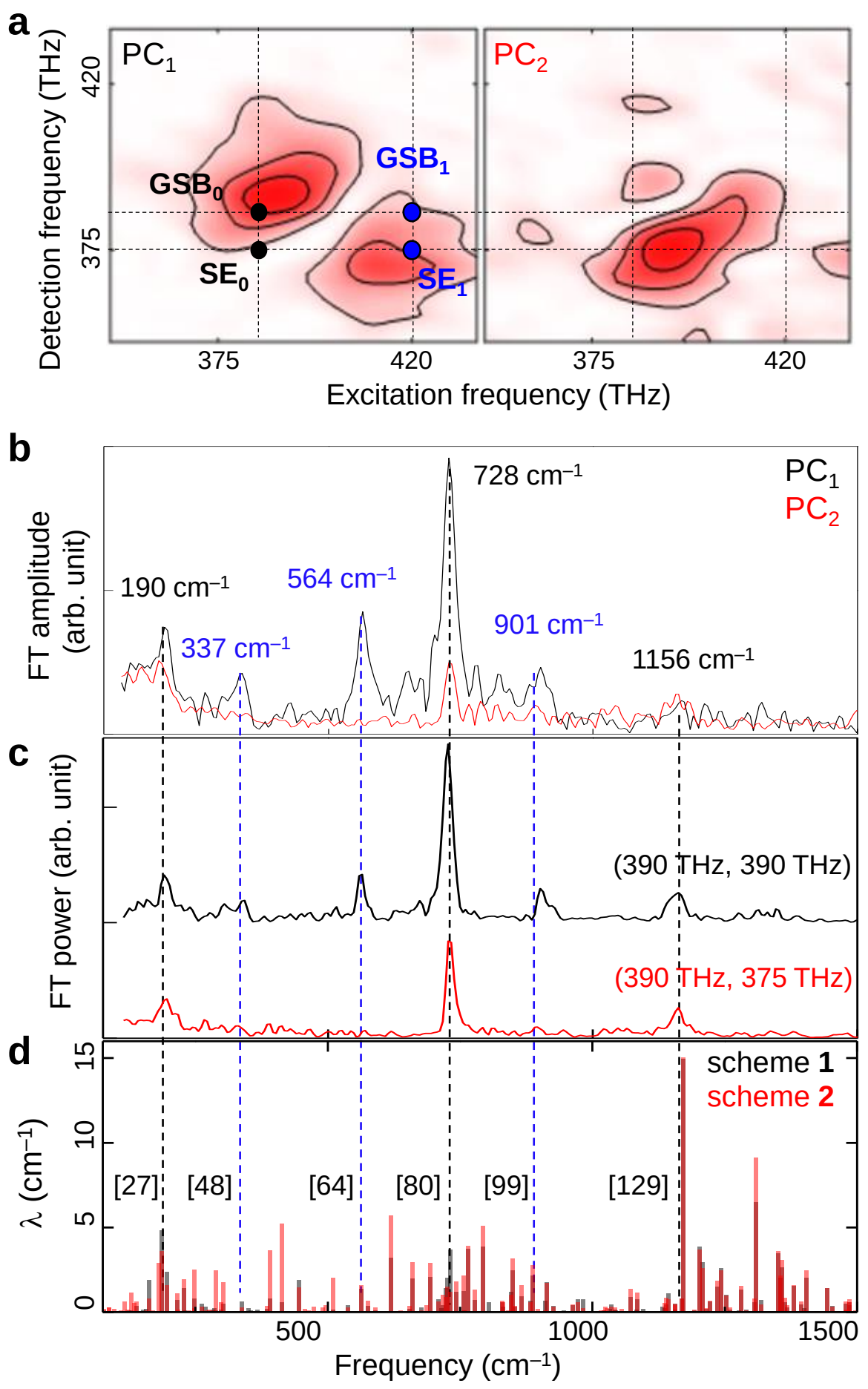

**Fig. 4 | Principal components analysis of BChla 2DES. a.** The 2D optical frequency spectra of two major principal components of the 2D coherent vibrational spectra for rephasing interaction pathway. **b.** The vibrational spectra of (**a**) multiplied by their score. Relatively intense six peaks at 190, 337, 564, 728, 901, and 1156 cm$^{-1}$ are marked with their centre frequencies. **c.** Coherent vibrational spectra obtained by Fourier-transforming the $T_w$-dependent 2DES signals at (390 THz, 390 THz) and (390 THz, 375 THz). The amplitudes of the three vibrational modes at 337,564, and 901 cm$^{-1}$ are significantly small in the CVS from the 2DES signal close to the $SE_0$ position. **d.** Theoretically calculated coherent vibrational spectra of BChla with respect to $Q_y$ excitation, where $\lambda$ indicates vibrational reorganization energy. The numbers in square brackets are the indices of the vibrational modes. See Supplementary Note 6 for more details on the simulation scheme **1** and **2**.

**Principal components analysis.** Vibrational coherence analysis at a single frequency point in the 2D map could cause a misinterpretation of the 2DES signal,[45] even though the relative amplitudes and linewidths of vibronically coupled modes found in the CVS obtained by Fourier-transforming the 2DES signal with respect to $T_w$ provide crucial information on their vibrational coupling strengths and decoherences.[46] Therefore, it would be useful to analyse both the coherent vibrational spectra and the 2D coherence maps to extract substantially detailed information on the nature and origin of coherent oscillations found in the spectrally congested $T_w$-resolved 2DES spectra. Here, we carried out a principal component analysis to identify the major 2D spectral patterns of the oscillating components in the 2DES signal and draw correlation between the vibronically coupled modes found in the CVS and the 2D spectra of major principal components. The entire three dimensional ($N_{\omega 1}\times N_{Tw}\times N_{\omega 2}$) 2DES data, where $N_{\omega 1}$, $N_{\omega 2}$, and $N_{Tw}$ are the numbers of data points of excitation frequency, detection frequency, and waiting time, respectively, are then represented with two 2D electronic spectra and two coherent vibrational spectra, which are the two principal components denoted as $PC_1$ and $PC_2$ in Fig. 4. All the other principal components are found to have negligibly small singular values.

Because the peak frequencies of the 2D coherence maps of the $PC_1$ and $PC_2$ overlap with the absorption and emission spectra of BChla, respectively, the majority of $PC_1$ can be identified with combination of $GSB_0$ and $SE_1$, and that of the $PC_2$ is $SE_0$ (see Supplementary Figure 6 for detailed descriptions of these peak assignments and for the corresponding PC spectra extracted from the non-rephasing 2DES signals). There are six strong peaks at 190, 337, 564, 728, 901, and 1156 cm$^{-1}$ in the coherent vibrational spectra associated with the two principal components, $PC_1$ and $PC_2$, of the 2DES data (Fig. 4b). The relative amplitudes of the three peaks at 337, 564, and 901 cm$^{-1}$ in the CVS of the $PC_2$ are found to be much weaker than those in the CVS of the $PC_1$. Such detection-frequency dependences of the three peaks are notable when comparing the single-frequency point CVS at the maximum positions of the 2D coherence maps of the $PC_1$ and $PC_2$ (Fig. 4c). Consistent with the present PCA results on the vibronically coupled modes, we find that the amplitudes of the 2D coherence maps at 337, 564, and 910 cm$^{-1}$ (Supplementary Figure 10) at (387 THz, 375 THz) are comparatively small. This indicates that the vibronic contributions from these modes to the $SE_0$ are negligible. The similarity and difference between the CVS of $PC_1$ and $PC_2$ suggest interesting information about the vibronic coupling strengths and their non-Condon effects, as will be discussed below.

**DFT calculation of vibronic coupling strength.** To assign the vibronically excited vibrational modes of BChla, we carried out two different density functional theory (DFT) calculations to estimate the vibronic coupling strengths. The first approach, scheme **1**, is to calculate the coordinate displacements of all the normal modes using the normal mode projection method,[46] which is based on direct comparisons of the optimized structures of the $S_0$ and $Q_y$ states (see the black bar spectrum in Fig. 4**d**). The second approach, scheme **2**, is based on the numerical calculation of the gradient of vibrational potential energy surface of the $Q_y$ state.[46] Once the slope of the vibrational potential energy curve at the equilibrium position of the ground state is obtained, one can estimate the vibrational coordinate displacement with invoking the harmonic approximation (red bar spectrum in Fig. 4**d**) (see Supplementary Note 6 for more details). Although these two approaches are not perfect, they still provide useful information about which modes are strongly coupled to the electronic $Q_y$ transition. From the DFT calculations, we could assign the six notable peaks in Fig. 4**b** at 190, 337, 564, 728, 901, and 1156 $cm^{-1}$ (see Supplementary Figure 11 for their vibrational eigenvectors). The peak frequencies of the observed vibrational modes of BChla are consistent with the previous experimental results obtained with hole-burning spectroscopy[47] and 2DES[23-25] studies.

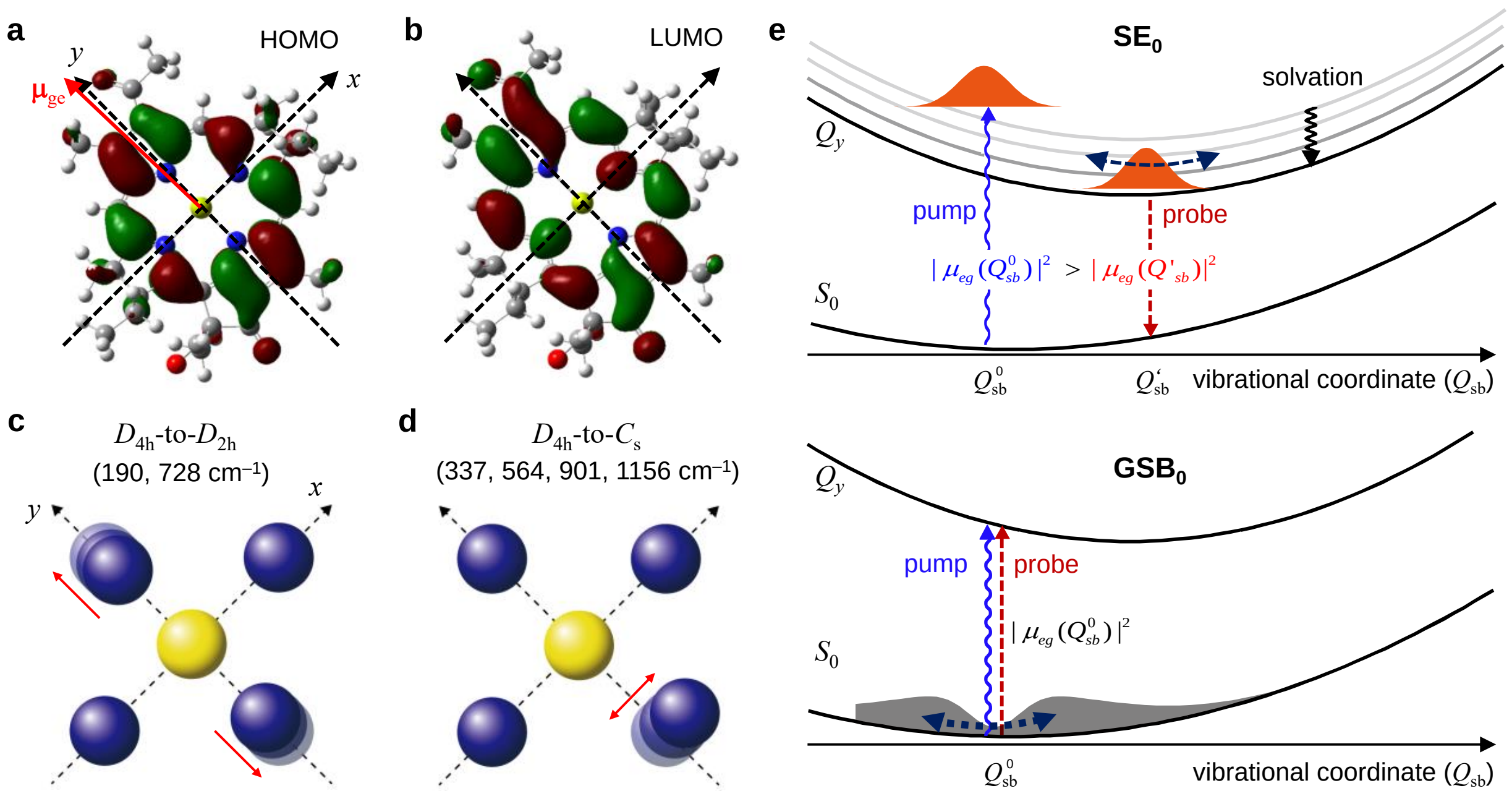

**Fig. 5 | Symmetry-breaking modes and non-Condon effect.** HOMO (**a**) and LUMO (**b**) of BChla. The thick red arrow represents the $Q_y$ transition dipole moment. ($\mu_{eg}$) Blue and yellow balls represent nitrogen and magnesium atoms, while the grey balls do carbon atoms. $x$- and $y$-axes are defined from two orthogonal N-Mg-N lines. Thin red arrows emphasizes the nuclear motions of N atoms in the BChla centre. The key eigenvector elements of the $D_{4h}$-to-$D_{2h}$ (**c**) and $D_{4h}$-to-$C_s$ (**d**) symmetry-breaking vibrational modes at the BChla centre are shown here. The full vibrational eigenvectors of these six modes are presented in Supplementary Figure 11. The red arrows in **c** and **d** show the directions (eigenvector elements) of nitrogen atoms. **e.** Schematic diagram describing a non-Condon effect on the reduced vibrational amplitudes of the three symmetry-breaking modes found in the CVS from the 2DES signal at (390 THz, 375 THz). Due to the destructive interference between the Franck-Condon factor and the Herzberg-Teller vibronic coupling, the oscillator strength associated with the $Q_{y,0}$-$S_0$ transition at the equilibrium coordinate $Q'_{sb}$ in the excited state is smaller than that at the equilibrium coordinate $Q^0_{sb}$ in the ground state. Such a non-Condon effect is also manifest in the broken mirror symmetry between the steady-state absorption and emission spectra in Fig. 3**a**.

**Non-Condon effects and symmetry.** The electronic structures of the HOMO and LUMO are shown in Figs. 5**a** and 5**b**. The transition dipole moment of the $Q_y$-transition is nearly parallel to the molecular $y$-axis ($\mu_{eg}$ in Fig. 5**a**). From them, it becomes clear that the electronic transition from the $S_0$ to $Q_y$ states includes a $D_4$-to-$C_{2h}$ symmetry-breaking element. This symmetry analysis explains why the two modes with frequencies of 190 and 728 cm$^{-1}$ that have $D_{4h}$-to-$D_{2h}$ symmetry-breaking characters are vibronically coupled to the $S_0$-$Q_y$ electronic transition (Fig. 5**c**). However, the vibrational motions of the other four modes with frequencies of 337, 564, 901, and 1156 $cm^{-1}$ induce a $D_{4h}$-to-$C_s$ symmetry-breaking (Fig. 5**d**).

Recently, the non-Condon effects on the excitonic energy transfer processes in photosynthetic systems have been studied and discussed in the theoretical and 2DES studies.[48-50] In general, non-Condon effects refer to the dependence of the transition dipole moment on the vibrational coordinates. Such a breakdown of Condon approximation is already manifest in the broken mirror symmetry between the steady-state absorption and emission spectra in Fig. 3**a**.[51] However, the linear spectra do not provide information about which molecular vibrations have strong Herzberg-Teller vibronic couplings. For those non-Condon modes, the transition dipole moment at the equilibrium position ($Q$') in the electronically excited state could differ from that at the equilibrium position ($Q^0$) in the ground state. In the 2DES experiments, such a non-Condon effect can be studied by examining the detection frequency-dependence of the CVS. As can be seen in Figs. 4**b** and 4**c**, the vibronic coupling strengths of the 337, 564, and 901 $cm^{-1}$ modes in the CVS associated with the $SE_0$ term are negligible, whereas they are strong in the CVS associated with the $GSB_0$ band. These disparities in the vibronic coupling strengths can be considered as experimental evidence of the non-Condon effects from these $D_{4h}$-to-$C_s$ symmetry-breaking modes. Here, it should be mentioned that the CVS of the GSB and SE contributions are similar to each other in the case of a dye (IR125) solution (Supplementary Figure 7). Therefore, the difference between the two CVS in Figs. 4**b** and 4**c** is a characteristic feature of BChla. To understand how such a difference in the coherent vibrational spectra is related to the breakdown of the Condon approximation, let us consider the transition dipole moment (TDM) associated with the probe beam interaction with BChla. The TDM determining the $SE_0$ signal should be evaluated at $Q$', whereas that associated with the $GSB_0$ is the TDM at $Q^0$ (Fig. 5**e**). In the present work, we performed all-parallel polarization 2DES measurements. However, it would be useful to carry out polarization-controlled 2DES spectroscopic studies, e.g., 2DES anisotropy measurements,[49] to extract more information about the relative directions of the absorptive and emissive transition dipole moments ($\mu_{eg}(Q^0)$ and $\mu_{eg}(Q')$ in Fig. 5e) in the future. Our experimental results suggest that the vibronic transition dipole moments are notably vibrational coordinate-dependent for the three (337, 564, and 901 cm$^{-1}$) modes, which cannot be revealed by linear spectroscopic studies of BChla. Noting that the molecular distortions along these three vibration coordinates ($Q_{sb}$) involve a $D_{4h}$-to-$C_s$ symmetry breaking, we anticipate that a further investigation about how such symmetry-breaking modes could affect the vibronic couplings and excitation transfer processes in photosynthetic reaction centers and light-harvesting complexes would be interesting.

## Discussion

In the present article, we showed that, as proof-of-principle experiments, direct time-domain measurement of the photon-echo response function of optical chromophores in solution is feasible using the SM-2DES technique. This measurement enables information on the correlation between two time-separated electronic coherences induced by a four-wave-mixing process and femtosecond-resolved molecular relaxation occurring on femtosecond to nanosecond time scales to be obtained with a single-point radio frequency detector and without a meter-long mechanical delay line. This work is the experimental demonstration of SM-2DES for studying the excited-state dynamics of a photosynthetic pigment (BChla) in solution. The efficient 2DES measurement of BChla solution within wide time and frequency windows by SM-2DES enabled observation and analysis of characteristic excited-state vibrational coherences of BChla. The experimental results on BChla can be of use to understand the nature of the photosynthetic energy transfer processes. We anticipate that SM-2DES will be useful for investigating molecular structure changes of chemically reactive systems due to its femtosecond time-resolvability and wide dynamic range measurement capability. In addition, the use of multiple single-point detectors would make polarization-dependent 2DES measurements efficient, which in turn could provide information on field-induced optical anisotropy of rotating molecules and electronic or vibrational optical activity of chiral molecules[52] in condensed phases. We believe that a few immediate applications of SM-2DES are not only to study excitonic dynamics in the manifold of one and multiple exciton states in solar harvesting complexes, semiconductors, and quantum dots but also to characterize the 2D optical response of chromophores in biological systems.

## Methods

**Synchronized mode-locked laser-based two-dimensional electronic spectroscopy.** Two Ti:Sapphire lasers (Rainbow 2, Femtolasers) producing a train of pulses with a duration time of 7 fs were used for the SM-2DES setup. The repetition frequencies of the MLs were phase-locked to a reference radio frequency (RF) signal from an 8-channel frequency synthesizer (Holzworth, HS9008A), the time-base of which is linked phase-coherently to a GPS-disciplined Rb atomic clock. The repetition frequencies of $ML_1$ and $ML_2$ were adjusted to be 80 MHz + $\Delta f_r$ and 80 MHz, respectively. The waiting time $T_w$ was automatically scanned due to the difference in the repetition frequencies, $\Delta f_r$, between the two MLs, which is the key aspect of the asynchronous optical sampling technique. The sampling rate was synchronized to the repetition rate of $ML_2$ (80 MHz) to record every pulse-to-pulse intensity difference of the signal. We set $\Delta f_r$ to be either 38.4 Hz or 3.2 kHz, which corresponds to $\Delta T_w$ values of 6 fs and 500 fs, respectively, for the SM-2DES measurements. Fractions of $ML_1$ and $ML_2$ are non-colinearly focused at a nonlinear crystal to generate a sum-frequency-generation field, which is utilized as an optical trigger for the automatic $T_w$-scan. The time jitter of the $T_w$-scan is determined by the pulse shape of the optical trigger and it was less than 1 fs. Due to the fast automatic scanning of $T_w$, $\tau_2$ can be scanned continuously by using a mechanical delay stage. The time interval for the $\tau_1$ step-scan was controlled to be $\Delta\tau_1$ = 1.868 fs, which corresponds to a translational stage step of 0.14 μm for each continuous $\tau_2$-scan. The time resolution of our SM-2DES was 12.5 fs, which was estimated by measuring the cross-correlation between $ML_1$ and $ML_2$ at the sample position. (see Supplementary Note 1 for details)

IR125 and BChla were purchased from Exciton and ChemCruz, respectively. The two molecules were used without further purification. IR125 (BChla) was dissolved in ethanol (1-propanol), and the light absorbance of the solution was adjusted to be approximately 0.2 (0.3), where the thickness of the sample cell is 200 μm. The thickness of the front window of the sample cell was as thin as 100 μm, which was to minimize additional pulse broadening by the optical cell itself. The solution sample continuously flowed with a gear pump (Micropump) during the SM-2DES measurements. The gear pump was cooled-down to 20 °C with a circulating chiller to protect the sample from heat generated from the gear pump. The ASOPS-based time-resolved spectroscopy inevitably wastes a lot of pulses because the automatic time scan involves significantly longer delays than electronic and vibrational relaxation times. Therefore, the duty cycle of such an ASOPS-based time-resolved vibrational spectroscopy of chromophores in the condensed phases could be pretty low. For example, if the time span is 5 ps for measuring vibrational dephasing constants and the pulse repetition period is 12.5 ns ($f_r$ = 80 MHz), the duty cycle[29] is estimated as 5 ps/12.5 ns, which is 0.04%. In this case, to protect the sample from photochemical and thermal damages caused by the interactions of the remaining 99.96% of pulses from each ML with the sample, it is necessary to block them by using a laser shutter synchronized with optical triggers. In contrast, a much longer scanning time (3.25 ns in this study) is required for the fast-scan-mode SM-2DES with $\Delta f_r$ = 3.2 kHz ($\Delta T_w$ = 0.5 ps) to characterize slow electronic relaxation dynamics. In this case, the duty cycle is as high as 26.0%. In order to minimize the optical damage on the sample, the incident beams were blocked by a set of optical shutters when data recording is not necessary. As a result of this gated sampling, the degradation rate of Bchla became three times slower. (see Supplementary Note 5 for more details) We checked that there was no additional signal from the sample degradation.

**Phase correction and amplitude calibration.** Small portions of the local oscillator fields from $ML_2$ before and after the $\tau_2$-scanning stage were taken using two beam splitters and were used to construct a Mach-Zehnder interferometer. This process provides the reference interferograms that are synchronized with the SM-2DES data collections. The reference interferograms are used to compensate for the phase fluctuation error in the detection frequency and the nonorthogonality between $T_w$- and $\tau_2$-scans.

In interferometric SM-PP spectroscopy, the probe field itself provides a phase reference automatically, and the spectral amplitudes of the measured PP spectrum can be simply normalized by using the pulse spectrum.[34] On the other hand, SM-2DES signals cannot be normalized, and their electric field phases cannot be easily corrected. However, the phase and amplitude of SM-2DES can be analytically calibrated and measured, respectively, by using interferometrically measured SM-PP spectra with exactly the same instrument (see Supplementary Note 3). The calibration factor is the ratio of the integrated SM-2DES spectrum over the excitation frequency to the interferometrically measured SM-PP spectrum. Unlike the conventional approach using the projection theorem, the present calibration and phase correction scheme works well because the complex (both real and imaginary) SM-PP signals can be obtained and used as both amplitude and phase references.

## Data availability

The data that support the findings of this study are available within the paper and its supplementary information files. Raw data are available from the corresponding author on a reasonable request.

## Acknowledgment

This work was supported by IBS-R023-D1.

## Author contributions

M.C., T.H.Y., and JW.K. conceived this experiment. JW.K. performed the experiments, J.J. developed a theory of SM-2DES. JW.K. performed the coherent vibrational spectroscopy simulation of BChla. JW.K., T.H.Y., and M.C. interpreted the experimental results. All the authors contributed to the writing of the manuscript.

## Competing interests

The authors declare no competing financial interest.

## Additional information

**Supplementary information** is available from the website.

Supplementary Information

# Two-dimensional electronic spectroscopy of bacteriochlorophyll *a* with synchronized dual mode-locked lasers

JunWoo Kim,[1] Jonggu Jeon,[1] Tai Hyun Yoon,[1,2*] and Minhaeng Cho[1,3*]

[1]*Center for Molecular Spectroscopy and Dynamics, Institute for Basic Science (IBS), Seoul 02841, Republic of Korea*

[2]*Department of Physics, Korea University, Seoul 02841, Republic of Korea*

[3]*Deparment of Chemistry, Korea University, Seoul 02841, Republic of Korea*

*Electronic mails: mcho@korea.ac.kr (MC) or thyoon@korea.ac.kr (THY)

**Supplementary Note 1**: Experimental methods

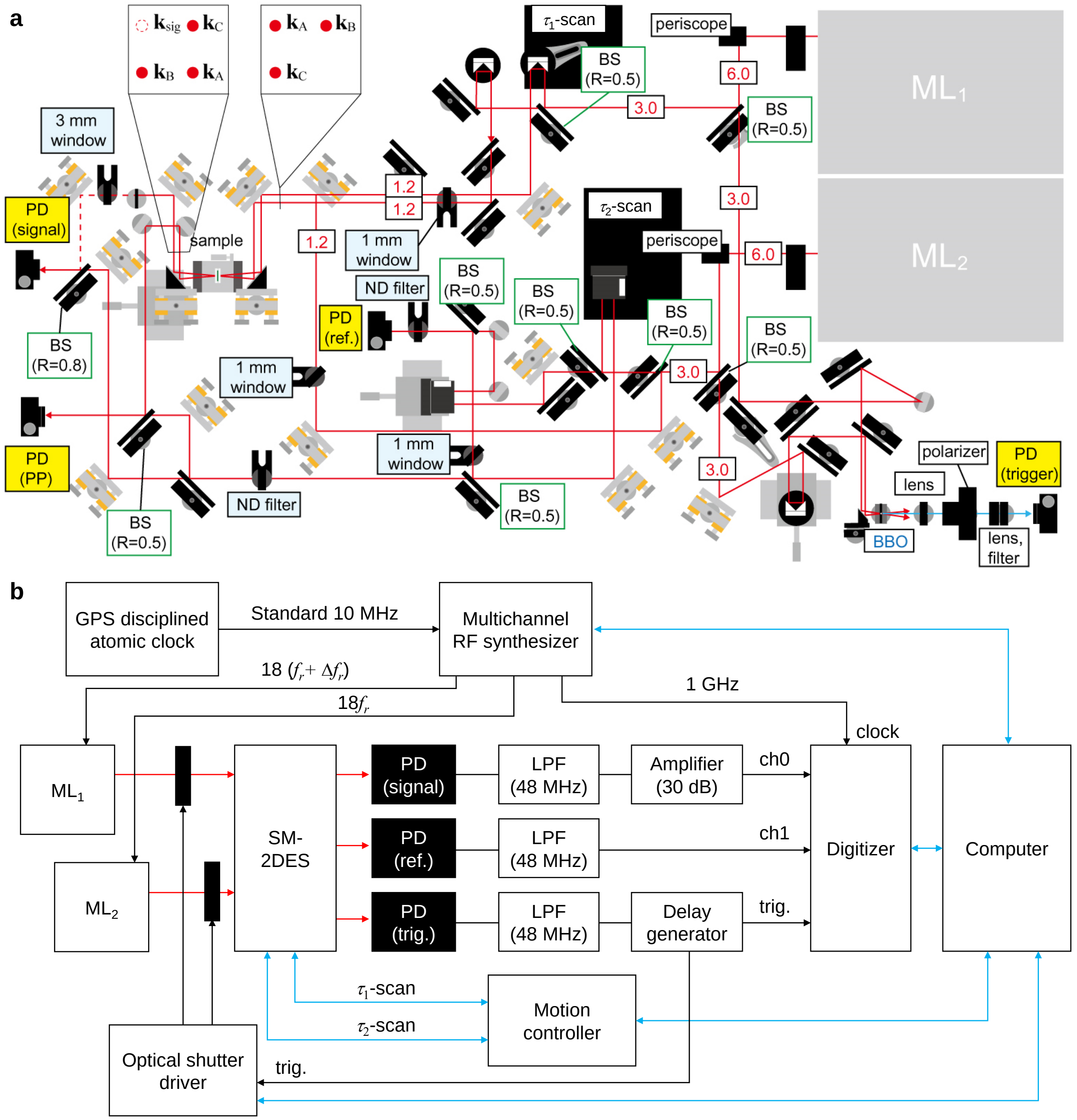

**Supplementary Figure 1| Experimental setup of SM-2DES. a**. Optical layout of SM-2DES. The red numbers on beamlines correspond to single pulse energies (in nJ) measured at these specific positions. The two boxes on the top-left corner show the front views of the optical beamlines before (right) and after (left) the sample. ML, repetition rate-stabilized mode-locked laser; BS, beam splitter; BBO, beta-barium borate crystal; PD, photodetector; ND filter, neutral density filter; PP: pump-probe. **b**. Schematic diagram of the SM-2DES instrumentation. Black (blue) arrows indicate unidirectional (bidirectional) electronic connections, while red arrows represent optical beam propagations. LPF, low-pass filter.

Supplementary Figure 1a depicts the laboratory optical layout of our SM-2DES system. Two 7-fs Ti:Sapphire mode-locked lasers (MLs; Rainbow 2, Femtolasers) were used as the light sources. The repetition rates of the MLs were phase-locked to the $18^{th}$ harmonics of 80 MHz + $\Delta f_r$ ($ML_1$) and 80 MHz ($ML_2$), respectively. Here, the ground state recovery time, which is usually the lifetime of the excited state, should be shorter than the pulse repetition period, $1/f_r$. Otherwise, undesired multi-photon processes via a train of pulses from high repetition rate lasers could make the SM-2DES signal obscure and the interpretation difficult. $\Delta f_r$ was controlled to be either 38.4 Hz or 3.2 kHz for femtosecond and nanosecond SM-2DES measurements, respectively, of the BChla solution. In the present work, we considered these $\Delta f_r$ values, 38.4 Hz and 3.2 kHz, to investigate two different relaxation dynamics that are the vibrational dephasing processes of vibronically coupled modes and the population relaxation of electronically excited states, respectively. The vibrational coherences dephase in a few picoseconds at room temperature while the population relaxation of electronically excited state takes place on the order of nanoseconds. To investigate them separately and more effectively, we performed SM-2DES measurements in two different time-scan modes, i.e., slow-scan mode with $\Delta T_w = 6$ fs ($\Delta f_r = 38.4$ Hz) for monitoring the underdamped oscillations of vibronically coupled modes and fast-scan mode with $\Delta T_w = 500$ fs ($\Delta f_r = 3.2$ kHz) for probing slow population relaxation processes. Half of the laser output was used for an optical trigger generation. For femtosecond-resolved measurements with $\Delta f_r = 38.4$ Hz, the separated beams for the optical trigger were non-collinearly focused at a 100-μm thick BBO crystal (CASTECH) with a parabolic mirror (f = 25 mm). The sum-frequency-generation signal produced by the temporal and spatial overlap of pulses from the two MLs is used as an optical trigger. For nanosecond measurements with $\Delta f_r = 3.2$ kHz, a 10-mm thick fused silica window was placed right before the parabolic mirror, and the 100-μm thick BBO crystal was replaced by a 1-mm thick BBO crystal. The pulse duration of the optical trigger must be larger than $\Delta T_w$ to avoid missing trigger pulse due to time-jitter.[1] The manual stage in the trigger line (the right-bottom side in Supplementary Figure 1a) sets the time-zero of the waiting time ($T_w = 0$).

Each pulse from $ML_1$ was further separated into two, and they are directed in the direction of $\mathbf{k}_A$ and $\mathbf{k}_B$. The optical time delay between them, denoted as $\tau_1$, was controlled by a motorized stage (XMS-160, Newport). The pulses from $ML_2$ propagate in the direction of $\mathbf{k}_C$. All of them were focused by a parabolic mirror with $f = 100$ mm and are collimated by another mirror with $f =$ 100 mm after the sample. Single pulse energy of each beam was measured to be about 1.2 nJ right before the sample. In the nanosecond SM-2DES measurements, the probe pulse energy was deliberately reduced to 0.6 nJ to minimize the undesired contributions from higher-order nonlinear optical signals and thermal heating effects. (see Supplementary Note 4 for details on the effect of local heating effect) The signal beam is generated along the direction of $\mathbf{k}_{sig} = -\mathbf{k}_A + \mathbf{k}_B + \mathbf{k}_C$ as a result of the third-order (in the electric fields) interaction at the sample. The front views of the four beams before and after the sample can be found in the boxes on the top-left corner of Supplementary Figure 1a. The signal electric field was combined with a local oscillator from $ML_2$, and the total intensity was detected at a PD (signal PD in Supplementary Figure 1a). The time delay between the signal and local oscillator denoted as $\tau_2$ was controlled by another motorized stage (XMS-160, Newport). Small portions of the local oscillator were separated before and after the $\tau_2$-scanning stage and used to construct a Mach-Zehnder interferometer used for phase referencing of each $\tau_2$-scan. The time-resolution of the experiment was estimated by means of measuring the transient grating signal from a BBO crystal that was placed at the sample position. A thin BBO crystal (thickness is less than 20 μm) is usually appropriate for characterizing the pulse width of a-few-fs light sources because of the broad sum-frequency-generation (SFG) bandwidth of such a thin nonlinear crystal. However, we found that the intensity of the sum-frequency field generated by $ML_1$ and $ML_2$ with a 20-μm-thick BBO crystal at the sample position is too weak to compress the optical pulse widths of the MLs efficiently. Therefore, we used a 100-μm-thick BBO crystal with a high SFG efficiency to measure the cross-correlation between $ML_1$ and $ML_2$. The estimated time resolution is 12.5 fs (full-width-at-half-maximum), even though the ideal time resolution for the present 2DES experiment when the pulse duration time is 7.0 fs would be 10.0 fs. This time resolution enables us to measure the coherent vibrational spectrum obtained from the Fourier transform of the SM-2DES data with respect to the waiting time. The upper limit of the measurable vibrational frequencies is 2665 $cm^{-1}$ because the corresponding oscillation period is 12.5 fs.

SM-2DES is advantageous for removing multiple artifacts occurring near $T_w = 0$. Firstly, it is nearly free from the solvent response and higher-order electronic response from the optical sample due to the low pulse energies (~1 nJ) involved in the experiment. The contribution from light scatterings by impurities, e.g., aggregates of solute molecules in the sample, does not affect the 2D spectra, because a full $T_w$-scan takes a time of $1/\Delta f_r$ while $\tau_1$ and $\tau_2$ are fixed at constants. Thus, the contribution from scatterings could be removed during the double Fourier transformations of $I(\tau_1, T_w, \tau_2)$ with respect to $\tau_1$ and $\tau_2$.

The data recording schematics of our SM-2DES is shown in Supplementary Figure 1b. A multichannel RF synthesizer (HS9008, Holzworth) phase-locked to a standard frequency provides reference frequencies to the two MLs. A digitizer with the bandwidth of 90 MHz (M3i.4861-exp, Spectrum) is synchronized to the standard frequency, and its sampling rate was set to be the same as the repetition rate of $ML_2$ (80 MHz). The trigger signal measured at the trigger PD passes a 48-MHz low-pass filter and enters a delay generator (DG645, Stanford Research Systems) as a trigger source. The delay generator triggers data acquisition and controls optical shutters. Two optical shutters are placed right after the MLs and are used to minimize the thermal damage on the sample. For femtosecond-resolved measurements with $\Delta f_r$ = 38.4 Hz, the shutters are synchronized to the optical trigger, so it is opened only at the detection time window. The same gated sampling cannot be achieved when we set the repetition frequency detuning factor $\Delta f_r$ to be 3.2 kHz because the $T_w$-scan rate ($\Delta f_r$ = 3.2 kHz) was faster than the bandwidth of the optical shutters (100 Hz). Instead, the shutters were opened when $\tau_2$ was being scanned.

The $\tau_1$- and $\tau_2$-scans were controlled by a motion controller (XPS-RL, Newport) with two high-resolution drivers (DRV-02L, Newport). The ranges of $\tau_1$- and $\tau_2$-scans were set large enough to fully record the interferogram envelope arising in the data (135 fs in optical time-delay). $\tau_2$ was scanned continuously with a constant velocity ($v_2$), which was stabilized with the 8 kHz servo rate by the motion controller. $v_2$ was set by considering the effective increment of $\tau_2$ ($\Delta\tau_2$) based on the relation

$$\Delta\tau_2 = \frac{2n_{air}v_2}{c\Delta f_r}, \tag{1}$$

where $n_{air}$ and $c$ are the refractive index of air and the speed of light, respectively. For femtosecond-resolved measurements with $\Delta f_r$ = 38.4 Hz, $\Delta\tau_2$ was set as large as possible, satisfying the Nyquist sampling condition, to achieve the highest data acquisition speed. The final $\Delta\tau_2$ value was 1.04 fs, which corresponds to $v_2$ = 5.0 μm/s. For nanosecond measurements with $\Delta f_r$ = 3.2 kHz, setting a large $\Delta\tau_2$ accompanying a high $v_2$ was inefficient because about 200 ms was required to stabilize the velocity at high $v_2$. Thus, the scan range became unnecessarily large to get larger $\Delta\tau_2$ at $\Delta f_r$ = 3.2 kHz. The optimized $\Delta\tau_2$ was 0.21 fs ($v_2$ = 50 μm/s) for the nanosecond-SM-2DES measurements. Finally, for every $\tau_2$-scan, the step of the mechanical delay stage used for $\tau_1$-scan was set to be 0.14 μm, which corresponds to 0.93 fs in the time domain. The interferometric pump-probe (IPP) data was recorded with the same instrument but without $\tau_1$-scan at the PD (see Supplementary Figure 1a).[1] The pump-probe data provides an absolute phase reference to the 2DES data (see Supplementary Note 3 for details). It should be mentioned that the size of a single $\tau_2$-scan data set ($N_{\tau 2} \times N_{Tw}$) is determined by the onboard memory of the fast digitizer, which is 2 GB with the resolution of 16 bit in the present work. Here, $N_{\tau 2}$, and $N_{Tw}$ are the numbers of $\tau_2$ and $T_w$ data points, respectively.

Although the absorption coefficient of BChla ($\varepsilon_{\mathrm{BChla}} \cong 5\times10^5$ $M^{-1}cm^{-1}$) is comparable to that of IR125 ($\varepsilon_{\mathrm{IR125}} \cong 10\times10^5$ $M^{-1}cm^{-1}$), the experimentally measured 2DES signal of BChla is found to be weaker than that of IR125. This discrepancy could result from the formation of a long-lived triplet state BChla.[2] Due to the high repetition rate of the pump ML used here, the population of the triplet state could be accumulated and reaches a quasi-equilibrium state during the course of measurements, which reduces the net population of the electronic ground state. Therefore, the SM-2DES signal amplitude could depend on the repetition rate of the ML. Currently, the effect of a high repetition frequency excitation beam is under investigation theoretically by numerically solving the kinetic equations involved with long-lived states and calculating the resulting 2DES signals upon varying repetition frequencies. (see Supplementary Note 3 for details)

The data acquisition time for a complete set of SM-2DES data is primarily determined by the velocity of $\tau_2$–scanning translation stage $v_2$ and the data transportation time from the digitizer to the computer storage. For femtosecond-resolved measurements with $\Delta f_r$ = 38.4 Hz, a single $\tau_2$–scan took about 5 s while the data acquisition took tens of milliseconds. In contrast, the data transportation time was the rate-determining process when we carried out nanosecond-SM-2DES measurements with $\Delta f_r$ = 3.2 kHz, where it took about 10 s in the present experiments. Thus, the total data acquisition times for femtosecond-resolved and nanosecond-SM-2DES with $\Delta f_r$ = 38.4 Hz and 3.2 kHz, respectively, were 7 min and 4 min. To improve statistics, we repeated such measurements 32 times for bacteriochlorophyll *a*, and report the average here.

**Supplementary Note 2**: Theoretical description of SM-2DES

In molecular spectroscopy, chromophores interact with incident electric fields and, within the electric dipole approximation, the radiation-matter interaction Hamiltonian can be written as

$$H_{\text{int}}(t) = -\boldsymbol{\mu} \cdot \boldsymbol{\mathcal{E}}(\mathbf{r},t), \quad (2)$$

where $\boldsymbol{\mu}$ is the electric dipole operator and $\boldsymbol{\mathcal{E}}(\mathbf{r},t)$ is the superposition of the electric fields used. Hereafter, we treat all the electric field quantities as scalar representing a common direction of polarization. The total Hamiltonian of the system is the sum of the system Hamiltonian $H_0$ in the absence of radiation and radiation-matter interaction Hamiltonian. The electric field of the pulse train $j$ can be written as

$$\mathcal{E}_j(\mathbf{r},t) = \frac{1}{2}\left[E_j(\mathbf{r},t) + \text{c.c.}\right], \quad (3)$$

where the complex electric fields are given by[3]

$$E_j(\mathbf{r},t) = \sum_{n=-\infty}^{\infty} A_j(t - n\Delta T_j) e^{i\left[\mathbf{k}_j(\omega_{cj})\cdot\mathbf{r} - \omega_{cj}(t - n\Delta T_j) - n\Delta\phi_{ceo,j}\right]} \quad (4)$$

in terms of the time interval $\Delta T_j$ between neighboring pulses of the $j$th optical frequency comb (OFC), the carrier frequency $\omega_{cj}$, and the carrier-envelope-phase shift $\Delta\phi_{ceo,j}$ between successive pulses.

For a coherent and periodic pulse train, the relative phase between two consecutive pulses in Supplementary Equation 4 must be an integer multiple of $2\pi$, and this imposes the following condition on the carrier frequency of the $j$th OFC laser,[3,4]

$$\omega_{cj} = n_{cj}\omega_{rj} + \omega_{\text{ceo},j}, \quad (5)$$

where $n_{cj}$ is the integer mode number characterizing the carrier frequency of the $j$th OFC. Here, we introduced the following definitions:

$$\begin{aligned} \omega_{rj} &= 2\pi/\Delta T_j, \\ \omega_{\text{ceo},j} &= \Delta\phi_{\text{ceo},j}\,\omega_{rj}/(2\pi). \end{aligned} \quad (6)$$

Then, $E_j(\mathbf{r},t)$ can be written as[4]

$$\begin{aligned} E_j(\mathbf{r},t) &= e^{i\mathbf{k}_j(\omega_{cj})\cdot\mathbf{r}-i\omega_{cj}t}\sum_{n=-\infty}^{\infty}A_j(t-n\Delta T_j) \\ &= e^{i\mathbf{k}_j(\omega_{cj})\cdot\mathbf{r}-i\omega_{cj}t}\sum_{n=-\infty}^{\infty}A_{nj}e^{-in\omega_{rj}t} \end{aligned}, \tag{7}$$

where $A_{nj}$ is the Fourier expansion coefficient of the pulse envelop function defined as

$$A_{nj} = (1/\Delta T_j)\int_{-\infty}^{\infty} A_j(t)e^{2\pi int/\Delta T_j}dt. \tag{8}$$

The OFC spectrum obtained by Fourier transforming Supplementary Equation 7 can be shown to have peaks at the following frequencies

$$\omega = \omega_{cj} + n\omega_{rj} = \omega_{\mathrm{ceo},j} + m\omega_{rj} \quad (m,n = \text{integers}), \tag{9}$$

where the second equality comes from Supplementary Equation 5. For later use, we introduce the differences in the repetition and carrier frequencies of the two OFC fields as

$$\begin{aligned} \delta\omega_r &= \omega_{r1} - \omega_{r2}, \\ \delta\omega_c &= \omega_{c1} - \omega_{c2}. \end{aligned} \tag{10}$$

In the present experiment, the sample, dye molecules in solution, interacts with three sets of pulse trains: (i) $OFC_1$ in the $\mathbf{k}_1$ direction, (ii) $OFC_1$ delayed by $\tau_1$ (which is much smaller than the repetition period of $OFC_1$ pulse train) from the original $OFC_1$ field and redirected in the $\mathbf{k}_2$ direction, (iii) $OFC_2$ in the $\mathbf{k}_3$ direction. The waiting time $T_w$ between the second and third interactions is scanned automatically by the pulse trains of $OFC_1$ and $OFC_2$ that have slightly different pulse repetition rates by $\delta\omega_r$. The signal in the $-\mathbf{k}_1 + \mathbf{k}_2 + \mathbf{k}_3$ direction is heterodyne-detected with the $OFC_2$ field as the local oscillator (LO), which is redirected in the signal direction and is delayed by $\tau_2$ (which is less than the repetition period of $OFC_2$ pulse train) from the third interaction. Note that $\tau_1$ and $\tau_2$ are explicitly controlled and scanned by, e.g., mechanical delay stages, but the waiting time $T_w$ is determined implicitly by the repetition frequencies of the two OFCs and their offset $\delta\omega_r$. Experimentally, the time origin for $T_w$ is set with an optical trigger signal produced when the sum-frequency-generation field produced by the overlap of the $OFC_1$ and $OFC_2$ pulses in a nonlinear crystal is detected.

The total electric field incident on the sample is the superposition of the three sets of pulse trains, i.e., $\mathcal{E}(\mathbf{r},t)=\mathcal{E}_1(\mathbf{r},t)+\mathcal{E}_2(\mathbf{r},t)+\mathcal{E}_3(\mathbf{r},t)$, where $\mathcal{E}_1(\mathbf{r},t)$ and $\mathcal{E}_2(\mathbf{r},t)$ are derived from the $OFC_1$ field and $\mathcal{E}_3(\mathbf{r},t)$ from the $OFC_2$ field. The three complex electric fields can then be written as follows

$$E_1(\mathbf{r},t)=e^{i\mathbf{k}_1\cdot\mathbf{r}-i\omega_{c1}(t+\tau_1)}\sum_{n=-\infty}^{\infty}A_{n1}e^{-in\omega_{r1}(t+\tau_1)},$$

$$E_2(\mathbf{r},t)=e^{i\mathbf{k}_2\cdot\mathbf{r}-i\omega_{c1}t}\sum_{n=-\infty}^{\infty}A_{n1}e^{-in\omega_{r1}t}, \tag{11}$$

$$E_3(\mathbf{r},t)=e^{i\mathbf{k}_3\cdot\mathbf{r}-i\omega_{c2}t}\sum_{n=-\infty}^{\infty}A_{n2}e^{-in\omega_{r2}t}.$$

Note that $t = 0$ corresponds to zero waiting time $T_w$, that is, zero time gap between the second and third field-matter interactions for infinitesimally narrow pulses. The third-order polarization generated by this electric field is given by

$$P^{(3)}(\mathbf{r},t)=\int_0^\infty dt_3\int_0^\infty dt_2\int_0^\infty dt_1 S^{(3)}(t_3,t_2,t_1)\mathcal{E}(\mathbf{r},t-t_3)\mathcal{E}(\mathbf{r},t-t_3-t_2)\mathcal{E}(\mathbf{r},t-t_3-t_2-t_1), \tag{12}$$

where $S^{(3)}(t_3,t_2,t_1)$ is the third-order response function. We then detect the signal along the direction of $\mathbf{k}_s = -\mathbf{k}_1 + \mathbf{k}_2 + \mathbf{k}_3$. The third-order polarization generating this signal can be written as $P^{(3)}(\mathbf{k}_s,t) = P^{(3)}(t)\exp(i\,\mathbf{k}_s\cdot\mathbf{r})$, which can be expressed as follows:

$$\begin{aligned}P^{(3)}(t)=&\frac{1}{8}\int_0^\infty dt_3\int_0^\infty dt_2\int_0^\infty dt_1 S^{(3)}(t_3,t_2,t_1)\\&\times\{E_1^*(t-t_3)E_2(t-t_3-t_2)E_3(t-t_3-t_2-t_1)\\&+E_1^*(t-t_3)E_3(t-t_3-t_2)E_2(t-t_3-t_2-t_1)\\&+E_2(t-t_3)E_1^*(t-t_3-t_2)E_3(t-t_3-t_2-t_1)\\&+E_2(t-t_3)E_3(t-t_3-t_2)E_1^*(t-t_3-t_2-t_1)\\&+E_3(t-t_3)E_1^*(t-t_3-t_2)E_2(t-t_3-t_2-t_1)\\&+E_3(t-t_3)E_2(t-t_3-t_2)E_1^*(t-t_3-t_2-t_1)\}\end{aligned}. \tag{13}$$

According to Supplementary Equation 11, within a time span $0 \le t < 2\pi/\omega_r$, $E_1$ always arrives first at the sample followed by $E_2$, then by $E_3$. On the other hand, the interaction times represented by the time arguments in each $E_k(\mathbf{r},t)$ in Supplementary Equation 13 do not necessarily follow this order. For example, the first term in the integrand corresponds to the situation where $E_3$ interacts first with the sample at time $t - t_3 - t_2 - t_1$, $E_2$ interacts next at $t - t_3 - t_2$, then $E_1$ at $t - t_3$, disregarding minor exceptions due to finite $\tau_1$ or $\tau_2$. From this consideration, only the sixth term above is consistent with the time ordering of the pulses, and all the other terms will only contribute to the signal only for very narrow time windows in $t_n$ ($n$ = 1, 2, 3) set by the pulse widths when two or more pulses overlap significantly. Effectively, each of these five terms becomes non-negligible only if one or more of $\tau_1$, $\tau_2$, or implicit waiting time $T_w$ is smaller than the pulse duration time and becomes a delta-function-like single point contribution in the impulsive limit. In particular, the first, third, and fourth terms in Supplementary Equation 13, where $E_3$ precedes $E_2$, would be non-negligible only for small $T_w$ (< pulse duration time), and therefore can be classified as coherent artifacts. In general, for a waiting time longer than the pulse width, these coherent artifacts would contribute negligibly. In the following development, we focus on the sixth term that is consistent with the time ordering of pulses and therefore is expected to make the predominant contribution to the observed signal. A complete theoretical investigation taking into account all six terms and numerical simulations will be reported elsewhere.[5]

Using the following notation for the $n$th peak frequency of the $j$th OFC counted from its carrier frequency

$$\omega_j^n = \omega_{cj} + n\omega_{rj}, \tag{14}$$

the sixth term in the integrand of Supplementary Equation 13 can be rewritten as

$$\begin{aligned} & E_3(\mathbf{r},t-t_3)E_2(\mathbf{r},t-t_3-t_2)E_1^*(\mathbf{r},t-t_3-t_2-t_1) \\ & = e^{i\mathbf{k}_s\cdot\mathbf{r}} \sum_{q,m,n=-\infty}^{\infty} A_{n1}^* A_{m1} A_{q2} e^{i(\omega_1^n-\omega_1^m-\omega_2^q)t} e^{i\omega_1^n\tau_1} e^{-i(\omega_1^n-\omega_1^m-\omega_2^q)t_3} e^{-i(\omega_1^n-\omega_1^m)t_2} e^{-i\omega_1^n t_1} \end{aligned}. \tag{15}$$

The polarization in $\mathbf{k}_s$ direction due to the sixth term is given by

$$\begin{aligned} P^{(3)}(\mathbf{k}_s,t) = \frac{1}{8} e^{i\mathbf{k}_s\cdot\mathbf{r}} \sum_{q,m,n=-\infty}^{\infty} A_{n1}^* A_{m1} A_{q2} e^{i(\omega_1^n-\omega_1^m-\omega_2^q)t} e^{i\omega_1^n\tau_1} \\ \times \tilde{S}^{(3)}(-\omega_1^n+\omega_1^m+\omega_2^q, \omega_1^m-\omega_1^n, -\omega_1^n) \end{aligned}, \tag{16}$$

where the frequency-domain response function (susceptibility) is defined as

$$\tilde{S}^{(3)}(\omega_3,\omega_2,\omega_1) \equiv \int_0^\infty dt_3 \int_0^\infty dt_2 \int_0^\infty dt_1 S^{(3)}(t_3,t_2,t_1) e^{i(\omega_3 t_3+\omega_2 t_2+\omega_1 t_1)} . \tag{17}$$

The electric field generated by the third-order polarization is[6]

$$E^{(3)}(\mathbf{k}_s,t) \propto \frac{i\omega_s}{n(\omega_s)} P^{(3)}(\mathbf{k}_s,t) , \tag{18}$$

where $n(\omega_s)$ is the refractive index of the medium at $\omega_s$. When the signal is heterodyne detected with the LO, which is the redirected and delayed $OFC_2$ field, the LO field can be written as

$$E_{\mathrm{LO}}(\mathbf{r},t) = e^{i\mathbf{k}_s\cdot\mathbf{r}-i\omega_{c2}(t-\tau_2)} \sum_{n=-\infty}^{\infty} A_{n2} e^{-in\omega_{r2}(t-\tau_2)} , \tag{19}$$

which is different from the original $OFC_2$ field $E_3(\mathbf{r},t)$ in Supplementary Equation 11 by the delay time $\tau_2$ and the propagation direction $\mathbf{k}_s$. The signal intensity at the detector is then given by the superposition of $E_{\mathrm{LO}}$ and the third-order signal field $E^{(3)}$ as follows

$$I(t) = \left|E_{\mathrm{LO}}(\mathbf{k}_s,t) + E^{(3)}(\mathbf{k}_s,t)\right|^2 = \left|E_{\mathrm{LO}}(\mathbf{k}_s,t)\right|^2 + \left|E^{(3)}(\mathbf{k}_s,t)\right|^2 + 2\,\mathrm{Re}\left[E^*_{\mathrm{LO}}(\mathbf{k}_s,t)E^{(3)}(\mathbf{k}_s,t)\right] . \tag{20}$$

In the present experiment, the LO intensity is comparable to that of the signal. Note that the first term on the right-hand side of Supplementary Equation 20 does not depend on $\tau_1$, the second term, a homodyne signal intensity, does not depend on $\tau_2$, and only the interference term does depend on both $\tau_1$ and $\tau_2$. Therefore, the first two terms can be easily removed by Fourier transforming the measured interferogram intensity with respect to $\tau_1$ and $\tau_2$. From Supplementary Equations 16-19, the interference signal can be written as

$$\begin{aligned} &2\,\mathrm{Re}\left[E^*_{\mathrm{LO}}(\mathbf{k}_s,t)E^{(3)}(\mathbf{k}_s,t)\right] \propto 2\,\mathrm{Im}[E^*_{\mathrm{LO}}(\mathbf{k}_s,t)P^{(3)}(\mathbf{k}_s,t)] \\ &= \frac{1}{4}\mathrm{Im}\left[\sum_{p,q,m,n=-\infty}^{\infty} A^*_{p2}A^*_{n1}A_{m1}A_{q2}\tilde{S}^{(3)}(-\omega_1^n+\omega_1^m+\omega_2^q,\omega_1^m-\omega_1^n,-\omega_1^n) e^{i(\omega_2^p+\omega_1^n-\omega_1^m-\omega_2^q)t} e^{i\omega_1^n\tau_1} e^{-i\omega_2^p\tau_2}\right] . \end{aligned} \tag{21}$$

The frequency factor associated with the measurement time $t$ in this equation can be rewritten as

$$\omega_2^p + \omega_1^n - \omega_1^m - \omega_2^q = -(m-n)\delta\omega_r + (p-q-m+n)\omega_{r2} . \tag{22}$$

In the experiment, only the slowly-oscillating interference terms in the RF domain are selectively detected with a low-pass filter or slow-response detector. Under this condition, the terms with non-zero coefficients of $\omega_{r2}$ vanish due to the relation $\omega_{r2} \approx 10^6\delta\omega_r$, and only the terms with the following values of $p$ and associated frequency factors survive when detecting the signal in time $t$

$$p = q + m - n, \quad \omega_2^p + \omega_1^n - \omega_1^m - \omega_2^q = -(m-n)\delta\omega_r\,. \tag{23}$$

The heterodyne-detected signal in Supplementary Equation 21 can then be written as

$$\begin{aligned}&2\,\mathrm{Re}\left[E_{\mathrm{LO}}^*(\mathbf{k}_s,t)E^{(3)}(\mathbf{k}_s,t)\right] \propto 2\,\mathrm{Im}[E_{\mathrm{LO}}^*(\mathbf{k}_s,t)P^{(3)}(\mathbf{k}_s,t)]\\&= \frac{1}{4}\mathrm{Im}\left[\sum_{q,m,n=-\infty}^{\infty} A_{(q+m-n)2}^* A_{n1}^* A_{m1} A_{q2} \tilde{S}^{(3)}(-\omega_1^n + \omega_1^m + \omega_2^q, \omega_1^m - \omega_1^n, -\omega_1^n) e^{i(n-m)\delta\omega_r t} e^{i\omega_1^n \tau_1} e^{-i\omega_2^{q+m-n}\tau_2}\right].\end{aligned} \tag{24}$$

Redefining the summation indices as $L = n - m$, $M = n$, and $N = q + m - n$, we have

$$\begin{aligned}&2\,\mathrm{Re}\left[E_{\mathrm{LO}}^*(\mathbf{k}_s,t)E^{(3)}(\mathbf{k}_s,t)\right] \propto 2\,\mathrm{Im}[E_{\mathrm{LO}}^*(\mathbf{k}_s,t)P^{(3)}(\mathbf{k}_s,t)]\\&= \frac{1}{4}\mathrm{Im}\left[\sum_{L,M,N=-\infty}^{\infty} A_{N2}^* A_{M1}^* A_{(M-L)1} A_{(N+L)2} \tilde{S}^{(3)}(-L\omega_{r1} + \omega_2^{L+N}, -L\omega_{r1}, -\omega_1^M) e^{iL\delta\omega_r t} e^{i\omega_1^M \tau_1} e^{-i\omega_2^N \tau_2}\right].\end{aligned} \tag{25}$$

We can express this signal in the following form:

$$\begin{aligned}&2\,\mathrm{Re}\left[E_{\mathrm{LO}}^*(\mathbf{k}_s,t)E^{(3)}(\mathbf{k}_s,t)\right] \propto 2\,\mathrm{Im}[E_{\mathrm{LO}}^*(\mathbf{k}_s,t)P^{(3)}(\mathbf{k}_s,t)]\\&= \frac{1}{4}\mathrm{Im}\left[\sum_{L,M,N=-\infty}^{\infty} B_{LMN} e^{iL\delta\omega_r t} e^{i\omega_1^M \tau_1} e^{-i\omega_2^N \tau_2}\right]\end{aligned} \tag{26}$$

with

$$\begin{aligned}B_{LMN} &= A_{N2}^* A_{M1}^* A_{(M-L)1} A_{(N+L)2} \tilde{S}^{(3)}(-L\omega_{r1} + \omega_2^{L+N}, -L\omega_{r1}, -\omega_1^M)\\&= A_{N2}^* A_{M1}^* A_{(M-L)1} A_{(N+L)2} \tilde{S}^{(3)}(\omega_{c2} + N\omega_{r2} - L\delta\omega_r, -L\omega_{r1}, -\omega_{c1} - M\omega_{r1})\end{aligned}. \tag{27}$$

This equation clearly shows that the signal under consideration that originates from the sixth term in Supplementary Equation 13 represents a rephasing pathway because the two coherence oscillation frequencies in the first and the third arguments of $\tilde{S}^{(3)}$ have opposite signs (disregarding non-zero comb indices). Using this result, we can rewrite the heterodyne-detected signal $I(\tau_2, t, \tau_1)$ as

$$I(\tau_2,t,\tau_1) = 2\,\mathrm{Re}\left[E^*_{\mathrm{LO}}(\mathbf{k}_s,t)E^{(3)}(\mathbf{k}_s,t)\right]$$
$$\propto \mathrm{Im}\left[\sum_{L,M,N=-\infty}^{\infty} B_{LMN} e^{iL\delta\omega_r t} e^{i\omega_1^M \tau_1} e^{-i\omega_2^N \tau_2}\right]. \tag{28}$$

To explore the relation of this signal with the waiting time $T_w$, we first introduce a complex function $S(\tau_2, t, \tau_1)$ defined as

$$S(\tau_2,t,\tau_1) = \sum_{L,M,N=-\infty}^{\infty} B_{LMN} e^{iL\delta\omega_r t} e^{i\omega_1^M \tau_1} e^{-i\omega_2^N \tau_2}\ , \tag{29}$$

which is related to the interference signal as $I(t_2, t, t_1) \propto \mathrm{Im}[S(t_2, t, t_1)]$. To a good approximation, the summation over $L$ in this equation can be replaced by the integral over a frequency variable $\omega$ that corresponds to $L\delta\omega_r$, and we obtain

$$S(\tau_2,t,\tau_1) = \frac{2\pi}{\delta\omega_r} \sum_{M,N=-\infty}^{\infty} \bar{\bar{B}}_{MN}(-t) e^{i\omega_1^M \tau_1} e^{-i\omega_2^N \tau_2}\ , \tag{30}$$

where $\bar{\bar{B}}_{MN}(t)$ is the inverse Fourier transform of $B_{LMN}$ regarded as a function of $\omega$ using the correspondence $\omega \leftrightarrow L\delta\omega_r$

$$\bar{\bar{B}}_{MN}(t) = \frac{1}{2\pi}\int_{-\infty}^{\infty} d\omega B_{(\omega/\delta\omega_r)MN} e^{-i\omega t}\ . \tag{31}$$

Using Supplementary Equation 27, $\bar{\bar{B}}_{MN}(t)$ can be shown to be related to the response function $\bar{S}^{(3)}(\omega_{c2}+N\omega_{r2},(\delta\omega_r/\omega_{r1})t,-\omega_{c1}-M\omega_{r1})$, which is a 2D Fourier transform of the time-domain response function $S^{(3)}(t_3, t_2, t_1)$ over $t_1$ and $t_3$, as a convolution with the pulse envelope function (see Supplementary Reference [4] for further details). Therefore, the signal detection time $t$ can be interpreted as the waiting time $T_w$ after down-scaling by the factor $\delta\omega_r/\omega_{r1}$. Now, the nonlinear response function that is related to the photon echo spectroscopy is denoted as $\mathrm{R}_{echo}(\tau_2, T_\mathrm{w}, \tau_1)$, and it is related to $S(\tau_2, t, \tau_1)$ as $\mathrm{R}_{echo}(\tau_2, T_\mathrm{w}, \tau_1) \propto iS(\tau_2, t, \tau_1)$. The definition of $\mathrm{R}_{echo}(\tau_2, T_\mathrm{w}, \tau_1)$ is given by a sum of two four-point time-correlation functions of electric dipole operators as (see Chapter 5 in Supplementary Reference 5 for details)

$$R_{echo}(t_3,t_2,t_1) = <\mu(0)\mu(t_1+t_2)\mu(t_1+t_2+t_3)\mu(t_1)\rho(-\infty)>$$
$$+<\mu(0)\mu(t_1)\mu(t_1+t_2+t_3)\mu(t_1+t_2)\rho(-\infty)>,$$

where $\rho(-\infty)$ is the density matrix at $t = -\infty$ and $\mu(t_1)$ is the electric dipole operator in the interaction picture. Using the correspondence between the waiting time and the measurement time, i.e., $T_w \leftrightarrow (\delta\omega_r/\omega_{r1})t$, we arrive at the following expression

$$I(\tau_2, \Delta t, \tau_1) \propto \mathrm{Re}[\mathbf{E}^*_{LO}(t) \cdot \mathbf{E}_{signal}(t)] \propto \mathrm{Re}[R_{echo}(\tau_2, T_w, \tau_1)]. \tag{32}$$

This equation is, theoretically, the principal result used to interpret the experimentally measured 2DES spectra in the main text.

**Supplementary Note 3**: Data analysis and phase correction

The raw SM-2DES data ($I(\tau_{2,q},T_{w,m},\tau_{1,n})$) measured and recorded in a series of time-averaged voltages at the signal PD contains the complete three-dimensional time-domain data. Since 2DES is useful for studying the correlation between the excitation and detection frequencies, the three-dimensionally restructured data, $I(\tau_{2,q},T_{w,m},\tau_{1,n})$, in the time domain should be Fourier transformed to $I(\omega_{2,q},T_{w,m},\omega_{1,n})$ in the frequency domain, where $\omega_1$ and $\omega_2$ are the Fourier conjugate frequencies associated with the two delay times $\tau_1$ and $\tau_2$, respectively. However, to obtain the correct 2DES spectra, appropriate time and phase references are needed. The time-zero of waiting time ($T_w$) was experimentally and automatically calibrated with each optical trigger (Supplementary Note 1). However, it should be noted that the relative positions of interference envelopes for each $\tau_2$-scan cannot be the same because the start times of the data acquisition and the continuous $\tau_2$-delay scan are not synchronized. The phase error caused by the fluctuation of envelope positions was compensated for using a reference interferogram that was measured at the reference PD (see Supplementary Figure 1a). The SM-2DES spectrum obtained by Fourier-transforming the measured time-domain signal with respect to $\tau_2$ becomes

$$I\left(\omega_{2,q'},T_{w,m},\tau_{1,n}\right)=\mathrm{FT}_{\tau_2}\left[I\left(\tau_{2,q},T_{w,m},\tau_{1,n}\right)\right]\exp[-i\Theta_{q',m,n}], \tag{33}$$

$$\text{where}\quad \Theta_{q',m,n}=\arg\left[\mathrm{FT}_{\tau_2}\left[I_{\mathrm{ref}}\left(\tau_{2,q},T_{w,m},\tau_{1,n}\right)\right]\right]. \tag{34}$$

Since the signal intensity $I(\tau_2,T_w,\tau_1)$ and the reference intensity $I_{\mathrm{ref}}(\tau_2,T_w,\tau_1)$ are phase-synchronized during the non-orthogonal $T_w$- and $\tau_2$-scans in the laboratory time, one can calibrate the phase of the signal with the phase of the reference signal (Supplementary Equation 34). We checked that the reference signal measured with a slow-response digitizer (< 1 MHz) could be of sufficient use for this phase calibration.

When $\mathbf{k}_A$ pulse precedes $\mathbf{k}_B$ pulse, the rephasing contribution to the 2DES signal is dominant, so that the rephasing signal ($I_{RE}$) corresponds to the raw data $I(\tau_{2,q},T_{w,m},\tau_{1,n})$ when $\tau_1$ is positive, i.e.,

$$I_{RE}\left(\omega_{2,q},T_{w,m},\tau_{1,n}\right)=I\left(\omega_{2,q},T_{w,m},\tau_{1,n}\right) \tag{35}$$

$$\text{with } \tau_n \geq 0 . \quad (36)$$

On the other hand, the non-rephasing contribution becomes important when $\mathbf{k}_B$ pulse interacts with the sample earlier than $\mathbf{k}_A$ pulse. In the conventional 2DES experiments, to obtain the non-rephasing signal separately, one of the mechanical translational stages was used to make $\tau_1$ negative. However, here we did not add the stage to our experimental setup for the sake of instrumental simplicity. Instead, the non-rephasing signal ($I_{NR}$) could be reconstructed with the following interpolation method,

$$I_{NR}\left(\omega_{2,q}, T_{w,m}, \tau_{1,n}\right) = \left(1 - r_n\right) I\left(\omega_{2,q}, T_{w,m}, \tau_{1,n}\right) + r_n I\left(\omega_{2,q}, T_{w,m+1}, \tau_{1,n}\right) \quad (37)$$

$$\text{with } r_n = \tau_n / \Delta T_w - \text{Int}\left[\tau_n / \Delta T_w\right], \text{and } \tau_n \leq 0 , \quad (38)$$

where Int[···] rounds the number inside the bracket to the nearest integer. Note that there should be a loss of time-resolution due to this interpolation scheme. However, this effect is really negligible when $\Delta T_w$ is sufficiently ultrashort compared to molecular relaxation processes of interest. The subscripts *RE* and *NR* will be omitted hereafter because the following data analyses will be performed for both rephasing and non-rephasing signals.

Finally, $I(\omega_2,T_w,\tau_1)$ is further Fourier transformed to $I(\omega_2,T_w,\omega_1)$ with respect to $\tau_1$ for both rephasing and non-rephasing signals. Due to the high repetition rate (80 MHz) of our synchronized ML system, the probe beam can be additionally diffracted by a slow thermal grating that is produced by repetitive nonradiative relaxations of electronically excited molecules by two non-collinearly propagating pump pulses (see Supplementary Note 4 for more details). The contribution from the transient thermal grating can be removed from the signal as follows:

$$I\left(\omega_{2,q}, T_{w,m}, \omega_{1,n}\right) = \text{FT}_{\tau_1}\left[I\left(\omega_{2,q}, T_{w,m}, \tau_{1,n}\right)\right] \cdot \exp\left[-i\theta_{q,n}\right] - R_{q,n} , \quad (39)$$

$$\text{with } R_{q,n} \exp\left[i\theta_{q,n}\right] = \left\langle \text{FT}_{\tau_1}\left[I\left(\omega_{2,q}, T_{w,m}, \tau_{1,n}\right)\right]\right\rangle , \quad (40)$$

where the average in Supplementary Equation 40 was taken by considering signals around $T_w = -500$ fs. Since the $ML_1$ and $ML_2$ pulses overlap in time every 12.5 ns, $T_w = -500$ fs means that $T_w = 12.5$ ns – 500 fs, which is sufficiently long enough to ignore any further electronic and vibrational relaxations, solvation dynamics, and rotational relaxation. Therefore, the experimentally measured signals at such long waiting times are almost constant in waiting time and purely originate from the thermal grating. Therefore, the average amplitude ($R_{q,n}$) and phase ($\theta_{q,n}$) of the thermal grating signal measured at around $T_w$ = –500 fs (Eq. **Error! Reference source not found.**) can be used to remove the thermal grating contribution from the raw data (see Eq. **Error! Reference source not found.**). The coherent vibrational spectra (Fig. 3e) extracted from the 2DES signals were obtained by the Fourier transform of $I(\omega_1,T_w,\omega_2)$ with respect to $T_w$, where the slowly decaying components associated with population relaxations of the electronically excited state, which could be fitted to a bi-exponential function, were removed from the raw data, $I(\omega_1,T_w,\omega_2)$, before the Fourier transformation with respect to $T_w$. We iteratively varied the apodization window to have a high signal-to-noise ratio and to minimize any loss of frequency resolution. We found that 2.5 ps is the optimum width of the apodization window, which corresponds to a frequency resolution of 8 cm$^{-1}$. Note that the data points at $T_w$ < 150 fs were not taken into account when we carried out Fourier transforms of the $T_w$-dependent 2DES signals, which was needed to remove the contribution from the coherent artifacts and the complexity due to the spectral overlap of GSB and SE terms at very short times ($T_w$ < 150 fs).

Hereafter, we focus on the phase-correction procedure. It is well-known that there exist two intrinsic problems in the conventional 2DES technique,[6] which are (i) the difficulty of calibrating the phase of the 2DES signal and (ii) the difficulty of correctly taking into consideration the effects of non-ideal (non-Gaussian) pulse spectra on the experimentally measured 2DES spectra. In most conventional 2DES with boxcar geometry, the phase of the complex 2DES spectrum was approximately calibrated by comparing its projected spectrum onto the probe frequency axis with the corresponding pump-probe spectrum measured independently. In stark contrast, not only the phase of the SM-2DES spectrum but also the probe pulse spectrum can be corrected by using the interferometrically measured pump-probe spectrum (IPS) with precisely the same experimental setup.[1]

Let us consider the IPS signal field of which spectrum is denoted as $E_{PP}(\omega_2, T_w)$ and 2DES field with its spectrum to be $E_{2D}(\omega_2, T_w, \omega_1)$. Unlike the conventional heterodyne-detected pump-probe spectroscopy, we could measure both the real and imaginary parts of $E_{PP}(\omega_2, T_w)$ simultaneously. Thus, we can use both the amplitude and phase spectra of $E_{PP}(\omega_2, T_w)$ to achieve the phasing correction of the 2DES spectrum. First, $E_{PP}(\omega_2, T_w)$ can be easily normalized using the probe spectrum ($E_0(\omega_2)$) measured independently. 2DES can be similarly expressed as a normalized form by introducing two complex scaling factors ($\alpha(\omega_2)$ and $\beta(\omega_2)$). $\alpha(\omega_2)E_0(\omega_2)$ represents the effective probe spectrum in the SM-2DES measurements. $\beta(\omega_2)$ is introduced here to correct the amplitude and phase of the 2DES signal. Thus, using the normalized forms of the IPS and the 2DES, the corresponding phase factors can be written as

$$\exp\left[i\Phi_{PP}\left(\omega_2,T_w\right)\right]=\frac{E_0\left(\omega_2\right)+E_{PP}\left(\omega_2,T_w\right)}{E_0\left(\omega_2\right)} \tag{41}$$

$$\exp\left[i\Phi_{2D}\left(\omega_2,T_w\right)\right]=\frac{\alpha\left(\omega_2\right)E_0\left(\omega_2\right)+\beta\left(\omega_2\right)E_{2D}\left(\omega_2,T_w\right)}{\alpha\left(\omega_2\right)E_0\left(\omega_2\right)}, \tag{42}$$

$$\text{where}\quad E_{2D}\left(\omega_2,T_w\right)\equiv\int d\omega_1 E_{2D}\left(\omega_2,T_w,\omega_1\right). \tag{43}$$

As can be seen in the definition of $E_{2D}(\omega_2, T_w)$, $E_{2D}(\omega_2, T_w)$ is the projected spectrum of $E_{2D}(\omega_2,T_w,\omega_1)$ onto the $\omega_2$ axis. The phase angle spectrum, $\Phi_{PP}(\omega_2,T_w)$, is an experimentally measurable quantity because both $E_0(\omega_2)$ and $E_{PP}(\omega_2, T_w)$ were experimentally measured. Also, the projected spectrum, $E_{2D}(\omega_2, T_w)$, is obtained from our experimentally measured $E_{2D}(\omega_2, T_w, \omega_1)$, which is the sum of the rephasing and non-rephasing 2DES signals when $T_w$ is sufficiently longer than the pulse duration time.[1] From the fact that the two phase-angle spectra, $\Phi_{PP}(\omega_2, T_w)$ and $\Phi_{2D}(\omega_2, T_w)$, must be the same, we have

$$\begin{aligned}\exp\left[i\Phi_{PP}\left(\omega_2,T_w\right)\right]-1&=\frac{\beta\left(\omega_2\right)E_{2D}\left(\omega_2,T_w\right)}{\alpha\left(\omega_2\right)E_0\left(\omega_2\right)}\\ &\equiv F\left(\omega_2\right)E_{2D}\left(\omega_2,T_w\right).\end{aligned} \tag{44}$$

The left-hand side of Supplementary Equation 44 is entirely determined using experimentally measured $E_0(\omega_2)$ and complex $E_{PP}(\omega_2, T_w)$. Also, we have $E_{2D}(\omega_2, T_w)$ on the right-hand side of Eq.**Error! Reference source not found.**. Thus, once $F(\omega_2)$ is obtained using Supplementary Equation 44, it can be directly used to achieve normalization of amplitude and correction of the phase of SM-2DES spectra.

In the present work, the two rate-determining steps in the data processing of SM-2DES are (1) reading the recorded data onto the data processing space and (2) removing the slowly decaying components from the time-domain data $I(\omega_1, T_w, \omega_2)$ by carrying out a nonlinear least-square fitting analysis. The latter was needed to carry out the Fourier transform of $I(\omega_1, T_w, \omega_2)$ with respect to $T_w$, which results in the coherent vibrational spectrum at $(\omega_1, \omega_2)$ in the 2D frequency space. It took about 4 min to average 48 data sets and about 1 min for Fourier-transforming the averaged data. However, we anticipate that such data analysis time can be significantly shortened by using a more efficient numerical analysis method in the future.

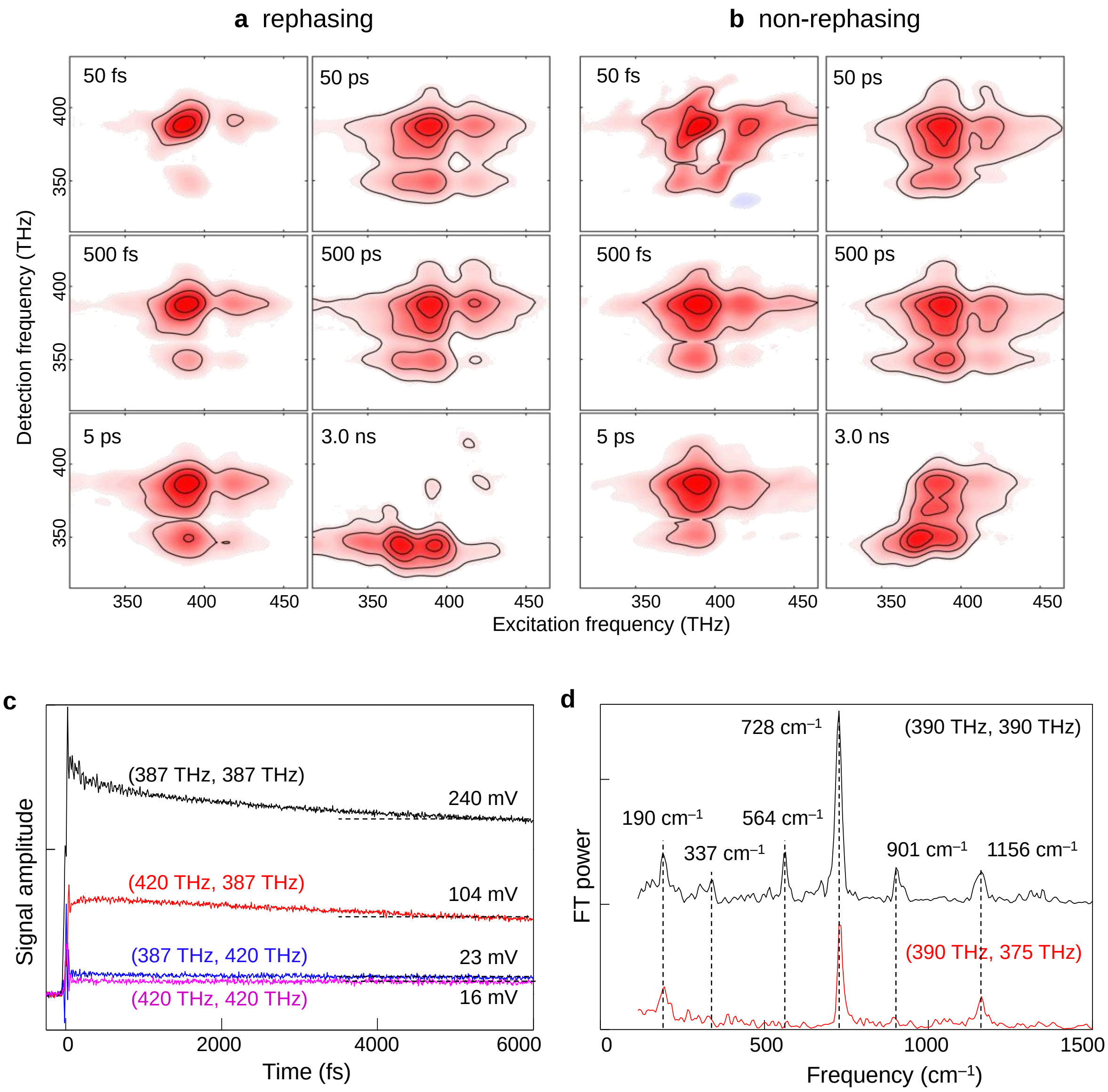

**Supplementary Figure 2| Supplementary 2DES data of BChla 1-propanol solution. a-b**, Rephasing (a) and non-rephasing (b) SM-2DES spectra of BChla 1-propanol solution at $T_w$ = 50 fs, 500 fs, and 5 ps ($\Delta f_r$ = 38.4 Hz), and at $T_w$ = 50 ps, 500 ps, and 3 ns ($\Delta f_r$ = 3.2 kHz). **c**, The total (rephasing + non-rephasing) absorptive 2DES time-profiles of BChla 1-propanol solution at (387 THz, 387 THz) (black, $GSB_0$ in the main text), (420 THz, 387 THz) (red, $GSB_1$ in the main text), (387 THz, 420 THz) (blue), and (420 THz, 420 THz) (violet). Each marked number in mV indicates the amplitude of each 2DES signal at 6000 fs. **d**, The coherent vibrational spectra extracted from the total (rephasing + non-rephasing) absorptive 2DES signals at (390 THz, 390 THz) (black), (390 THz, 375 THz) (red), where the two points are the maximum amplitude points of the 2D spectra of $PC_1$ and $PC_2$ in Fig. 4. It should be emphasized that the time step of 6 fs and the scanning time up to 6 ps for the $T_w$-scan data in S2c, obtained with using an ASOPS scheme, are very accurate because the automatic time-delay generation is controlled by locking the repetition frequencies to the atomic clock-generated reference frequency. Therefore, the frequencies of vibronically coupled modes (in Fig. R5 and S2d) do not require any frequency calibration.

**Supplementary Note 4**: Thermal grating effect

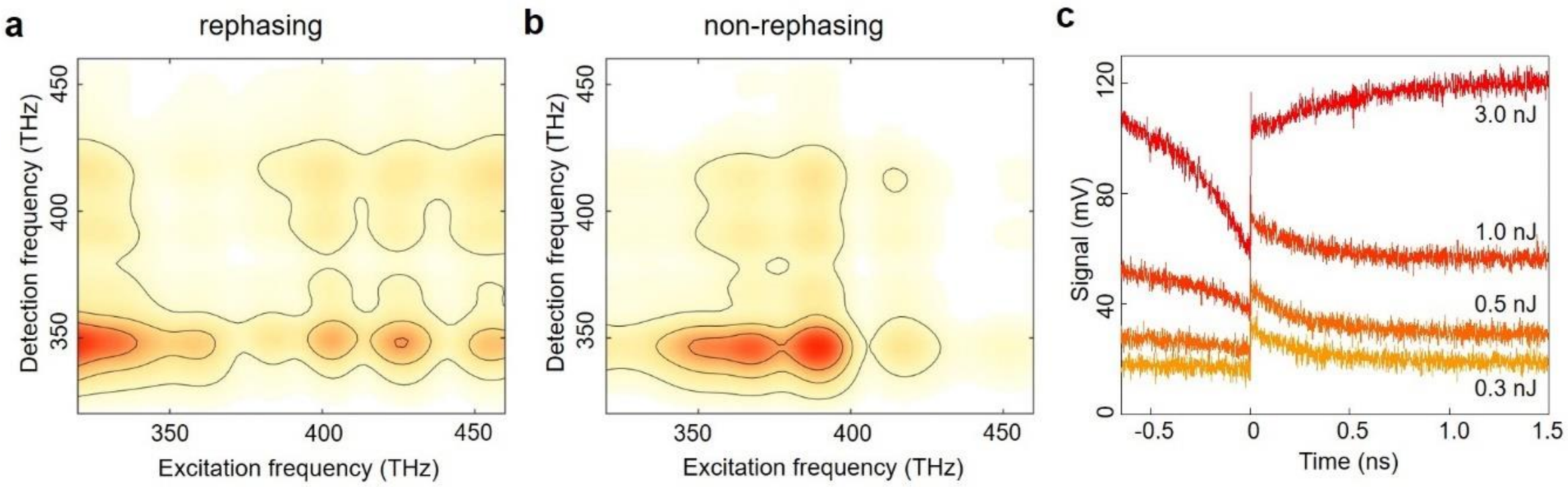

**Supplementary Figure 3| Contribution of the thermal grating to the SM-2DES signal. a**,(**b**,) The rephasing (non-rephasing) SM-2DES signal of IR125 ethanol solution at $T_w < 0$ with the probe pulse energy of 0.5 nJ. **c**. The $T_w$-scan data at $\tau_1 = 0$ for varying probe pulse energy.

In 2DES experiments, transient grating generated by an interferometric excitation of chromophores by two non-collinearly propagating pump pulses disappears in time due to population relaxation of the excited state. The electronic transition energy is then converted into heat, increasing local temperature around chromophores. If such an excitation-dissipation process is repeated rapidly, the system won't have sufficient time to relax to a thermal equilibrium state. In the present SM-2DES, we use high repetition rate MLs ($f_r$ = 80 MHz). Due to the non-collinear geometry, where the two pump and one probe pulses all propagate in different directions, a thermal grating is created, which then undesirably contributes to the 2DES spectrum. In fact, such thermal grating contribution was ignored entirely in the conventional 2DES experiments mainly because the Ti:Sapphire laser used has a repetition rate of 1 kHz, so it was safe to assume that the thermal grating is fully relaxed and disappeared before the next pump interaction after 1 ms. This is not the case for our SM-2DES utilizing 80 MHz MLs. If the thermal diffusion is much slower than the repetition rate, the system will reach a new quasi-equilibrium state, so the thermal grating does not interfere with the 2DES data, and instead, it provides a phase reference as described in Supplementary Note 3. Supplementary Figure 2a and b show the rephasing and non-rephasing SM-2DES signal at $T_w$ = 12.5 ns –500 fs, respectively, where the thermal grating contribution is predominant. In particular, the broad spectral features along both excitation and detection frequencies of the two 2D spectra indicate that they originate from an electronically non-resonant response of the sample. However, if the population relaxation of a given photochemical system is dominated by nonradiative processes, the SM-2DES can be of limited use because the thermal grating formed could diffract the incident light beams.

To further confirm that the thermal grating contribution to the SM-2DES signal cannot be ignored when high repetition rate lasers with sufficiently strong power, we additionally carried out 2DES measurements for varying probe pulse energy (Supplementary Figure 3c). In principle, both the thermal grating and the electronically resonant response signals should increase together with the probe intensity. Indeed when the probe pulse energy is relatively low, e.g., 0.3 and 0.5 nJ, both the background level and the decay component in the $T_w$-scan data, which correspond to the thermal grating and the transient grating signal, respectively, linearly depend on the probe pulse energy. However, if the probe pulse energy is high, e.g., 1.0 and 3.0 nJ, the background signal at around $T_w = 0$ shows a negative dip, which makes the numerical analysis of data difficult. Since the pump pulse energies are the same at a fixed value when we obtain the four different data shown in Supplementary Figure 3c, the negative dip at around $T_w = 0$ when the probe energy is high cannot originate from the difference in pump pulse energy. Thus, the unexpected and undesired negative feature in the background signal arises from a probe-induced effect. To minimize it, we lowered the probe power down to 0.6 nJ when we carried out SM-2DES measurements with $\Delta T_w$ = 500 fs. Fortunately, even with slightly high power probe pulses (1.2 nJ), the background signal remains constant for almost several picoseconds. Therefore, when we carried out SM-2DES measurements with $\Delta T_w = 6$ fs, we used probe pulses whose energy was 1.2 nJ.

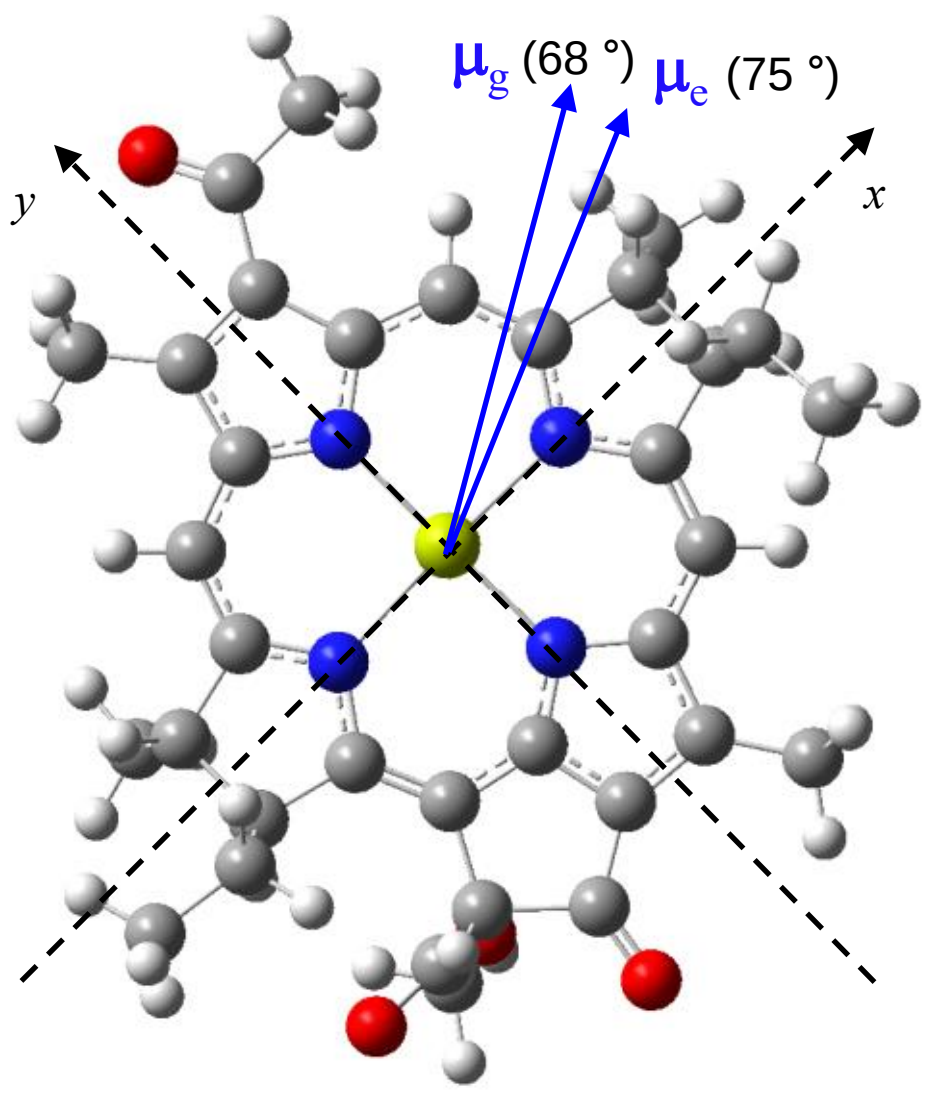

**Supplementary Figure 4| Electric dipole moment of BChla.** The electric dipole moments of BChla in $S_0$ ($\mu_g$) and $Q_y$ ($\mu_e$) states are represented by blue arrows. The numbers in parentheses are the angles between the *y*-axis and the electric dipole moments.

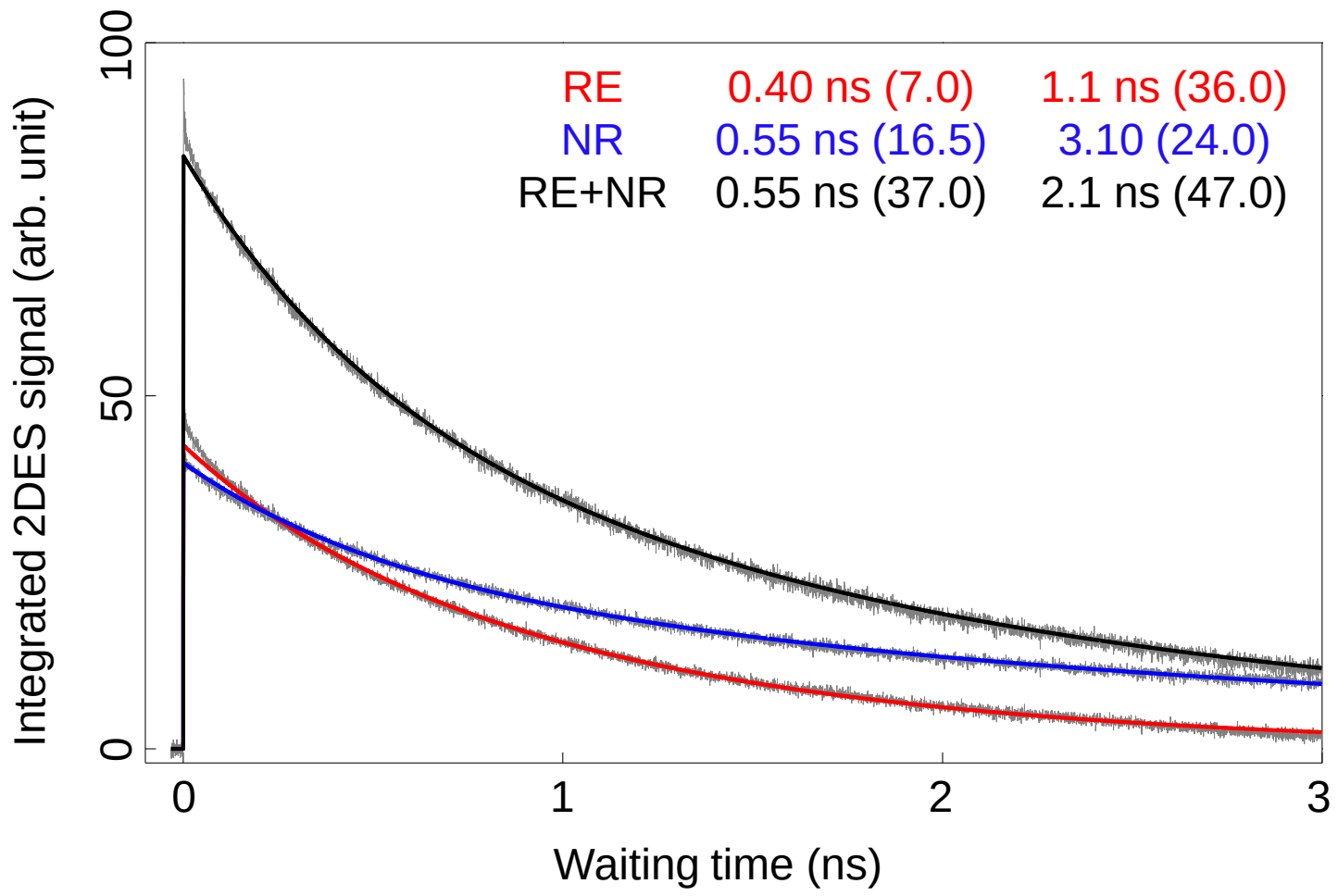

**Supplementary Figure 5| Slow decay components in BChla.** The grey curves represent the integrated rephasing (RE), non-rephasing (NR), and total (RE+NR) 2DES signals. The bi-exponential fitting results of RE (red), NR (blue), and total (black) signals are overlaid on the raw data. The fitting parameters are given in the upper part of this figure, where the numbers in the parentheses represent the amplitude of the relevant decay components. The fitting window in this nonlinear least-square fitting is from 0.25 ns to 3.0 ns.

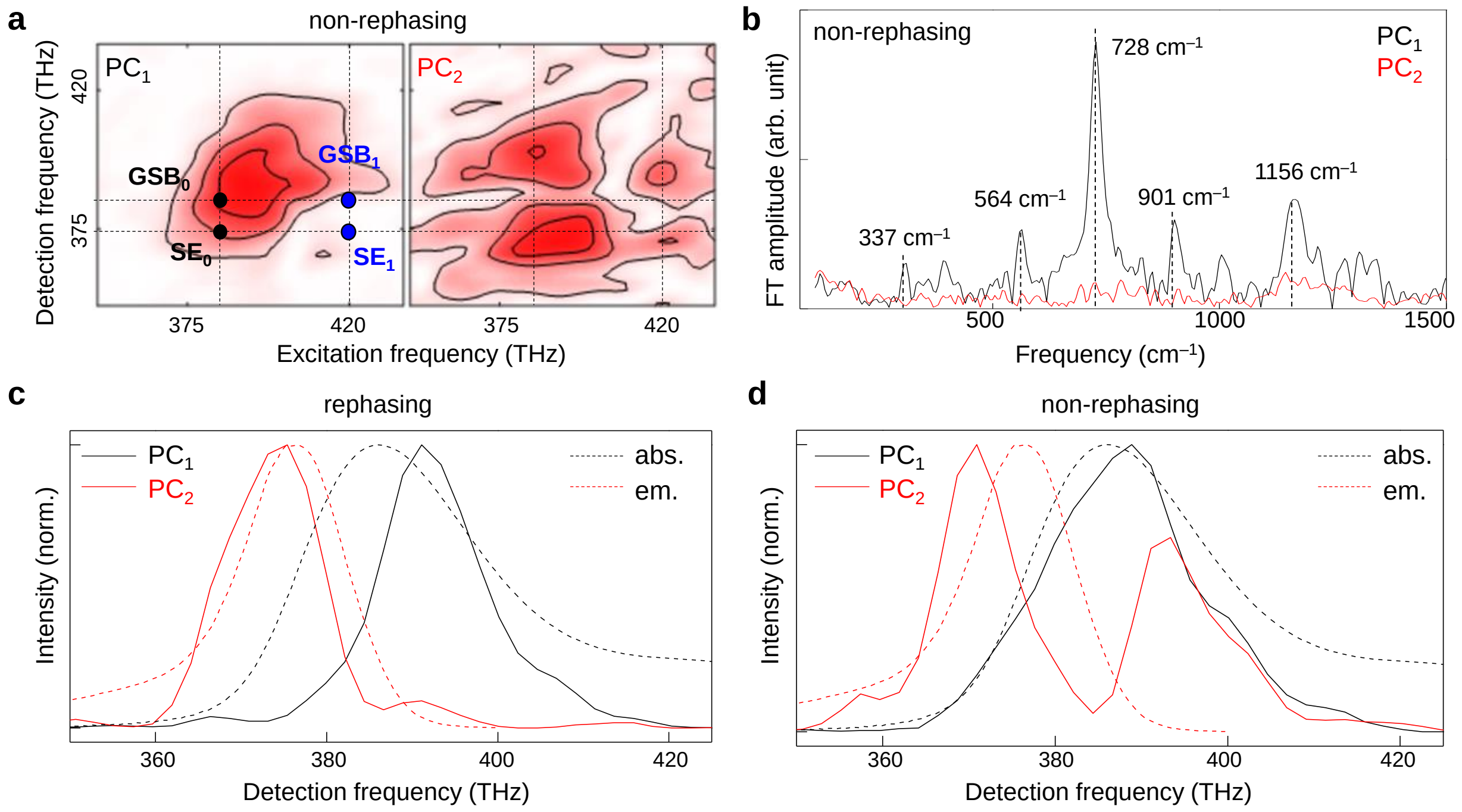

**Supplementary Figure 6| Non-rephasing 2D coherent vibrational spectra of BChla. a,** The 2D optical frequency spectra of two major principal components ($PC_1$ and $PC_2$) of the 2D coherent vibrational spectra for non-rephasing interaction pathway. **b,** The vibrational spectra of (**a**) multiplied by their score. Relatively intense five peaks at 337, 564, 728, 901, and 1156 $cm^{-1}$ are marked with their centre frequencies. **c-d,** The 2D electronic spectra of $PC_1$ (black) and $PC_2$ (red) integrated over excitation frequency (< 400 THz) for rephasing (**c**) and non-rephasing (**d**) interaction pathways. The absorption (black, dashed line) and emission (red dashed line) spectra of BChla 1-propanol solution are overlaid for comparison.

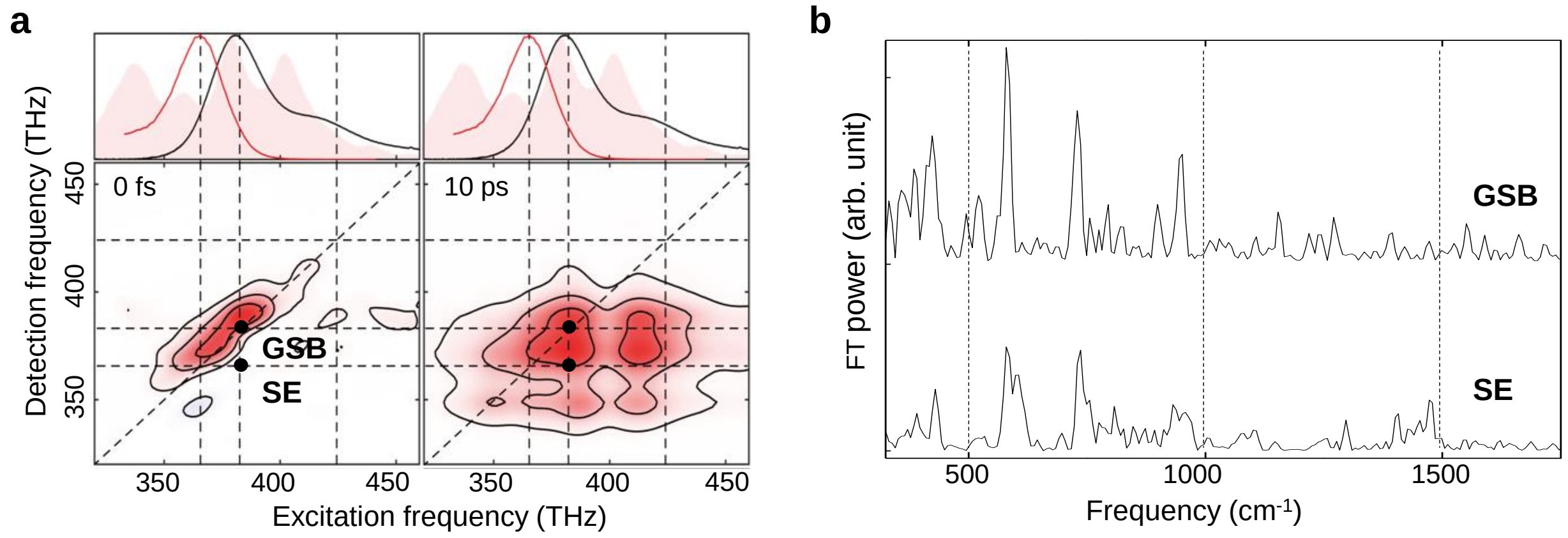

**Supplementary Figure 7| Coherent vibrational spectrum of IR125/ethanol solution. a.** Rephasing SM-2DES spectra of IR125/ethanol solution at $T_w = 0$ ($\Delta f_r = 32$ Hz) and 10 ps ($\Delta f_r = 1.6$ kHz). The absorption (black) and emission (red) spectra of IR125 in ethanol are shown in the top panels. In addition, the power spectrum of the light source (pink area) is presented for the sake of direct comparison. **b**. Coherent vibrational spectra of IR125 at the stimulated emission (SE) and ground-state bleach (GSB), which are marked by black circles in (**a**).

**Supplementary Note 5**: Gated sampling to protect samples from photochemical damage

Chlorophyll molecules, including BChla, are easily oxidized in the atmosphere and damaged by the photo-induced local heat. In conventional 2DES experiments with low repetition rate lasers, the chlorophyll solution is degassed before the measurement, and its 2DES signal is measured with a rotating cell for room temperature measurement. However, the rotating cell, which usually provides hundreds of rotations per second, is not recommended for SM-2DES with high-repetition-rate MLs due to its slow refresh rate.

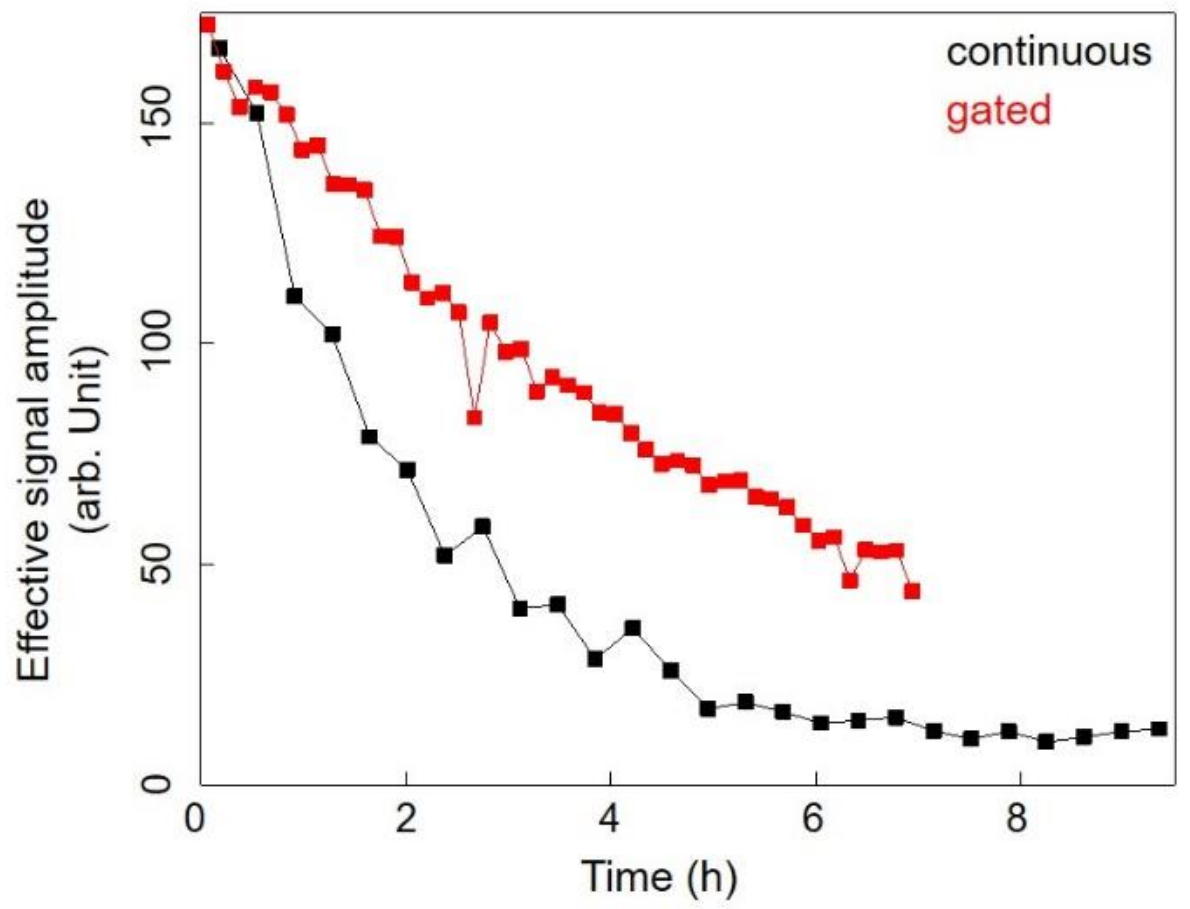

**Supplementary Figure 8| The effect of gated sampling for the SM-2DES measurement of BChla/1-propanol solution at room temperature.** The effective amplitude of the 2DES signal with continuous (black) and gated (red) sampling as a function of measurement time.

In this research, we flowed the BChla solution with a gear pump, which provides a pulsation-less liquid circulation, and gated sampling described in Supplementary Note 1 to minimize the photo-thermal damage of BChla molecules. We also chilled the gear pump with a circulating chiller to reduce the heat generated at the pump. The SM-2DES signal disappears in 5 min under the absence of the pump-chilling system. Supplementary Figure 3 shows the amplitude of the SM-2DES signal as a function of laboratory time. Since the decay of the 2DES signal corresponds to the BChla degradation, it can be confirmed that the use of gated sampling (red in Supplementary Figure 8) extends the lifetime of BChla by three times. Although it is impossible to remove the photochemical damage and to prevent the sample reservoir from being exposed to air, the SM-2DES can measure tens of data set before the degradation effect becomes dominant. Among the 46 data sets measured with the gated sampling, only 33 data sets were averaged because we lose signal-to-noise ration with more than 33 data sets.

**Supplementary Note 6**: Coherent vibrational spectrum simulation

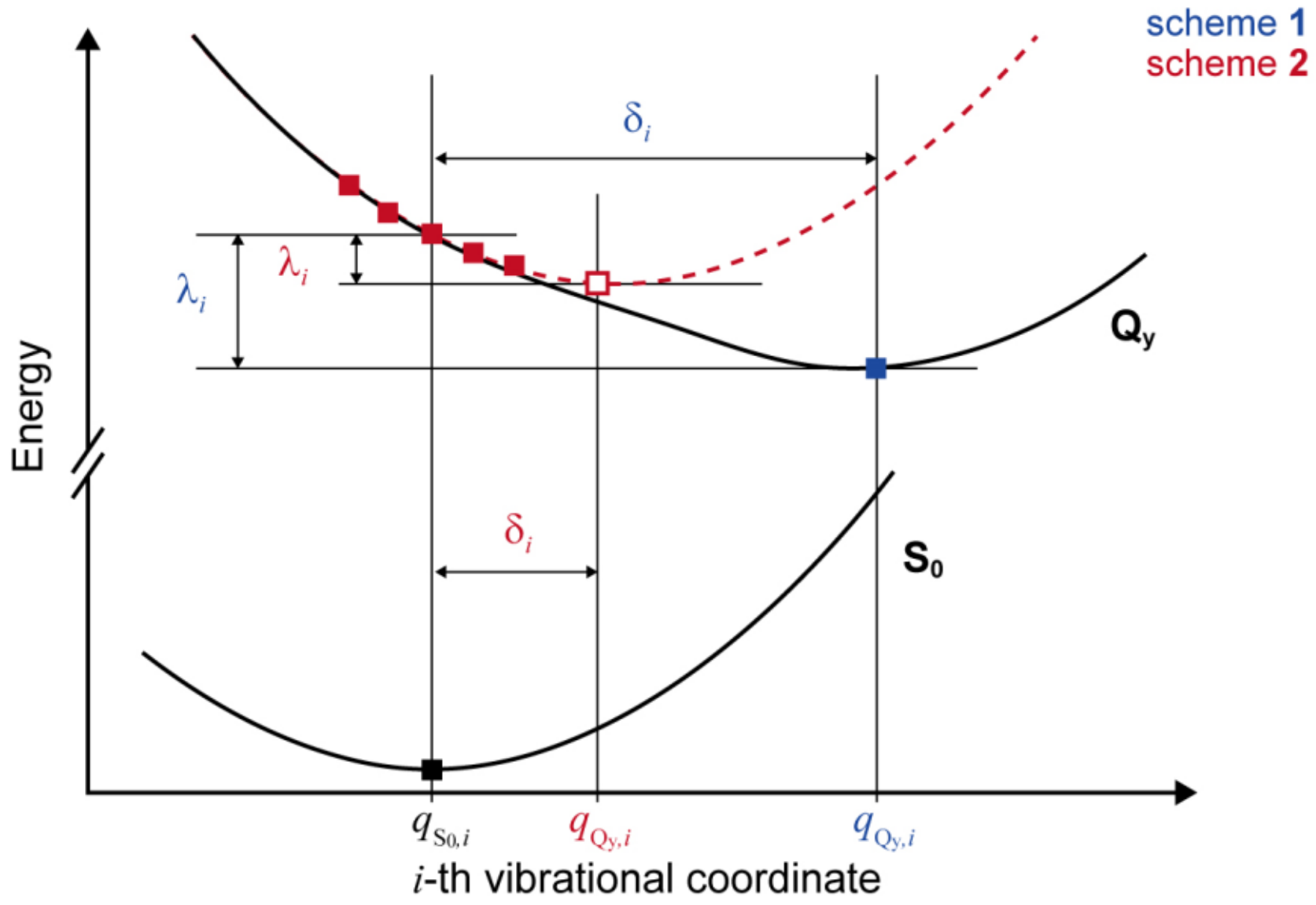

**Supplementary Figure 9| Coherent vibrational spectrum simulation.** The schematic diagram for the two coherent vibrational spectrum simulation schemes. $q_{\mathrm{A},i}$ indicates the equilibrium position of the electronic state "A" of the *i*th vibrational coordinate. $\delta_i$ and $\lambda_i$ are the displacement and vibrational reorganization energy of *i*-th mode, respectively. Blue (red) coloured parameters are utilized for scheme **1** (scheme **2**) introduced in the main text. The filled squares indicate the quantum mechanically calculated points, while the empty square does the point estimated by extrapolation in scheme **2**.

The amplitude of a coherently oscillating vibrational mode activated by an impulsive photoexcitation, which corresponds to the vibronic coupling strength of the vibrational mode, is determined by the potential energy surface (PES) difference between the electronic ground and excited states involved in the relevant transition. Within the harmonic approximation with respect to the PES along all vibrational coordinate of the molecular system, the coupling strength of each vibrational mode can be simply estimated by calculating the geometrical displacement between the two electronic states along the vibrational coordinate. Therefore, once we know the equilibrium structure of the ground state ($\boldsymbol{q}_{\mathrm{S0}}$) and its vibrational mode, it is possible to estimate the initial time-evolution of the excited molecule. In this research, we employed two types of simulations (scheme **1** and **2** in the main text) to assign the vibrational peaks in the coherent vibrational spectrum (CVS), which is the Fourier transform of the nuclear wave packet oscillation observed in time-resolved spectroscopy, observed in the SM-2DES data of BChla.

The two schemes require the geometrical difference between the equilibrium structures of the $S_0$ and $Q_y$ states in terms of vibrational coordinates because the amplitude of each mode in CVS is proportional to its vibrational reorganization energy ($\lambda_i = h\nu\delta_i^2/2$). Supplementary Figure 8 illustrates how the two schemes calculate the vibrational displacement ($\delta_i$) between two equilibrium positions ($\boldsymbol{q}_{S0,i}$ and $\boldsymbol{q}_{Qy,i}$) on a vibrational coordinate. The difference between the two schemes is how to estimate the equilibrium structure of $Q_y$ state ($\boldsymbol{q}_{Qy}$). In scheme **1**, $\boldsymbol{q}_{Qy}$ is directly calculated with a quantum chemical calculation and represents the structural difference as the linear combination of vibrational coordinates using the normal-mode projection method.[7] In scheme **2**, we calculated five single-point energies of the $Q_y$ state by varying the displacement of the vibrational modes of the $S_0$ state. Based on the five points, the minimum energy structure of the $Q_y$ state can be estimated by solving the quadratic equation.

If the actual PESs of a molecule satisfy the harmonic approximation, the calculation results from schemes **1** and **2** should be the same. In general, however, the harmonic approximation is only applicable for the initial state ($S_0$ in this research), but not for the final state ($Q_y$ in this research) as illustrated in Supplementary Figure 9. In case the equilibrium structural difference between the two electronic states is significant, scheme **1** will provide more accurate information on the excited state dynamics, while scheme **2** is appropriate for the negligible difference. Although we employed these methods just for peak assignment due to the large size of BChla, these methods are quite accurate enough for small chromophores to investigate the molecular dynamics of the excited state.[8]

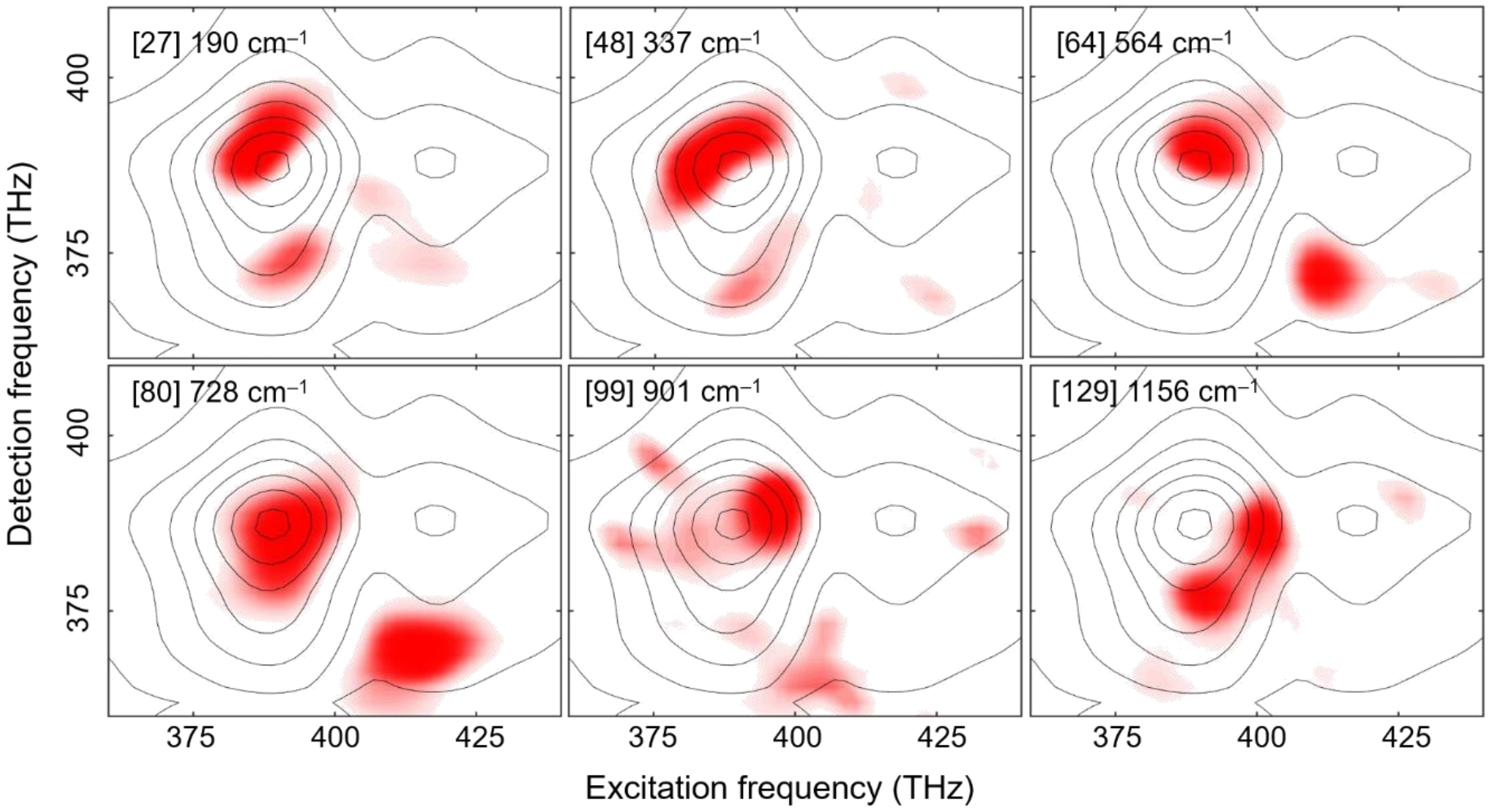

**Supplementary Figure 10| Rephasing 2D vibrational coherence maps of BChla.** 2D vibrational coherence maps of BChla at 190, 337, 564, 728, 901, and 1156 cm$^{-1}$ extracted from the corresponding rephasing 2DES data. The number in each square bracket indicates the normal mode index, and the contour line represents the rephasing 2DES spectrum at 2 ps.

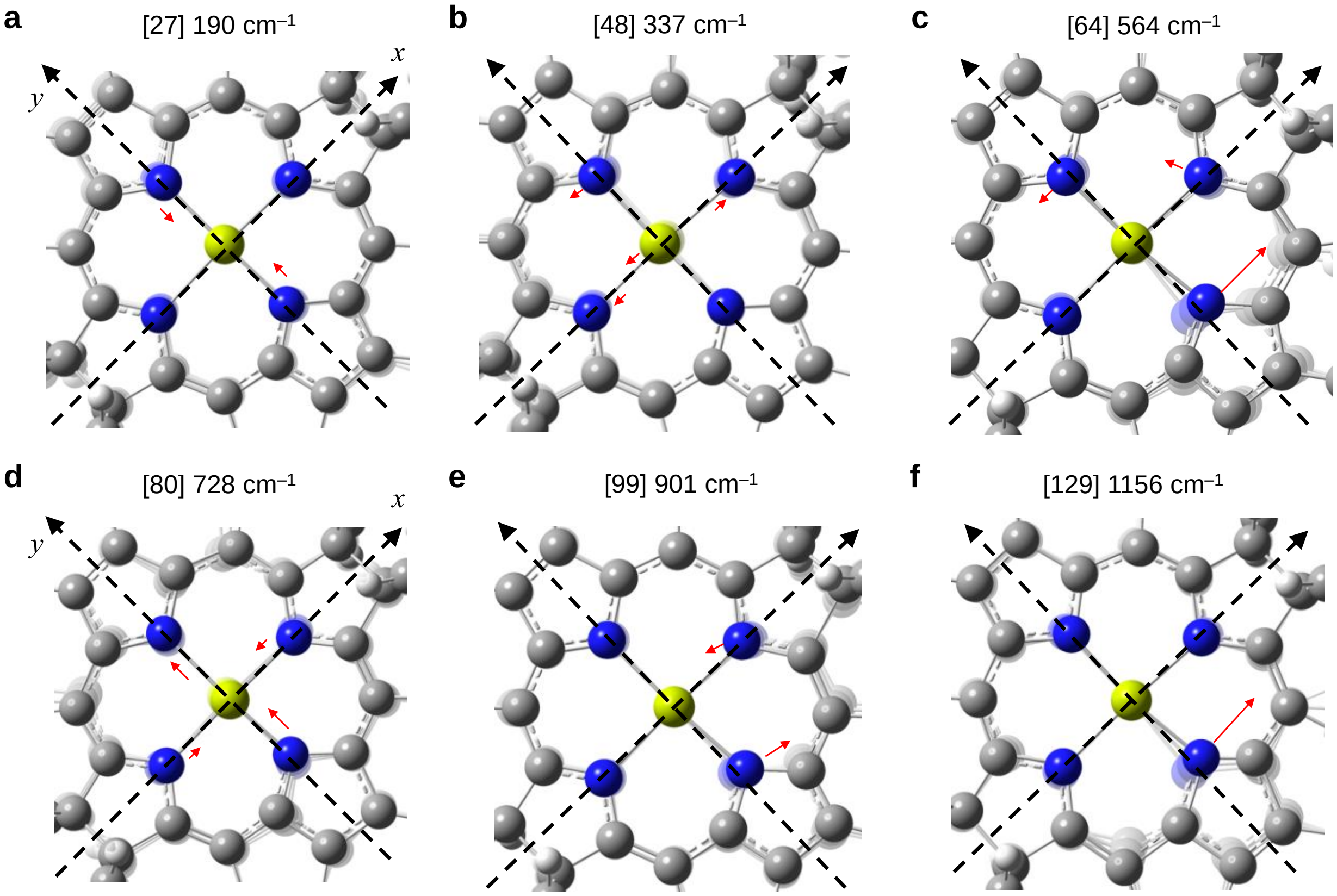

**Supplementary Figure 11| Vibrational eigenvectors of six modes that are strongly coupled to the $Q_y$ transition.** The number in each square bracket indicates the normal mode index. The two dashed arrows are the molecular *x*- and *y*-axes (see Fig. 5). The red arrows represent the eigenvector elements.

## SUPPLEMENTARY REFERENCES